\documentclass{aastex}
\usepackage{emulateapj5}

\def\lesssim{\mathrel{\hbox{\rlap{\hbox{\lower4pt\hbox{$\sim$}}}\hbox{$<$}}}}
\def\gtrsim{\mathrel{\hbox{\rlap{\hbox{\lower4pt\hbox{$\sim$}}}\hbox{$>$}}}}

\shorttitle{Binary Formation in Magnetic Clouds}
\shortauthors{Nakamura \& Li}

\begin{document}

\title{Binary and Multiple Star Formation in Magnetic Clouds:
Bar Growth and Fragmentation}
\author{Fumitaka Nakamura}
\affil{Faculty of Education and Human Sciences, Niigata University,
8050 Ikarashi-2, Niigata 950-2181, Japan; fnakamur@ed.niigata-u.ac.jp}
\and
\author{Zhi-Yun Li}
\affil{Department of Astronomy, University of Virginia, P.O. Box 3818,
Charlottesville, VA 22903; zl4h@virginia.edu}

\begin{abstract}
In the standard scenario of isolated low-mass star formation, strongly 
magnetized molecular clouds are envisioned to condense gradually 
into dense cores, driven by ambipolar diffusion. Once the cores 
become magnetically supercritical, they collapse to form stars. 
Previous studies based on this scenario are limited to axisymmetric 
calculations leading to single supercritical core formation.
The assumption of axisymmetry has precluded a detailed investigation 
of cloud fragmentation, generally thought to be a necessary step 
in the formation of binary and multiple stars. In a series of papers, 
we studied the non-axisymmetric evolution of initially
magnetically subcritical clouds, using a two-dimensional
magnetohydrodynamic code based 
on the physically motivated thin-disk approximation. 
We found that such clouds become unstable to non-axisymmetric perturbations
after the supercritical cores are formed due to ambipolar
diffusion.  In this paper, we focus on the evolution of clouds 
perturbed by an $m=2$ mode of a modest fractional amplitude of
5\%, with an eye on binary and multiple star formation. We show 
that for a wide range of initial cloud parameters, the $m=2$ mode 
grows nonlinearly into a bar during the isothermal collapse 
after the supercritical core formation. The instability is
driven by the domination of the magnetically-diluted gravity 
over the combined thermal and magnetic pressure gradient in
the supercritical cores. Such gravity-dominated cores can break 
up into fragments during or after the isothermal phase of cloud 
evolution. 

The outcome of fragmentation depends on the initial cloud conditions,
such as the magnetic field strength, rotation rate, amount of cloud 
mass (relative to thermal Jeans mass) and mass distribution. It is 
classified into three different types: (1) {\it separate core formation},
in which the bar ($m=2$) mode breaks up into two separate cores 
during the isothermal collapse, with a core separation of order
$10^4$ AU, (2) {\it bar fragmentation}, in which the $m=2$ mode evolves 
into a needle-like, opaque ``first bar'' (at a density $n\gtrsim 10^{12}$ 
cm$^{-3}$), which breaks up into multiple fragments with initial
masses of order $10^{-2}$ M$_\odot$ and separations of order 
$10^2$-$10^3$ AU, and (3) {\it disk fragmentation}, in which the bar growth 
remains slow during the isothermal collapse and the central region 
evolves into a rapidly rotating, opaque ``first disk'', which 
breaks up into several self-gravitating blobs with separations 
less than the disk size ($\lesssim 10^2$ AU). These three types 
of fragmentation loosely correspond to the empirical classification
of embedded binary and multiple systems of Looney, Mundy, \& Welch,
based on millimeter dust continuum observations. The well-studied 
starless core, L1544, appears to belong to the bar fragmentation 
type. We expect it to produce a highly elongated, opaque bar at 
the center in the future, which should break up into fragments of 
initial masses in the substellar regime. 
\end{abstract}


\section{Introduction}
\label{sec:introduction}

Most main-sequence stars are found in binary or low-order 
multiple-star systems \citep{Duquennoy91, DFischer92}.
Recent observations of star-forming regions have revealed that
the binary frequency of pre-main sequence stars is as high as or even higher 
than that of the main-sequence stars 
\citep{CLeinert93,AGhez97,RKohler00,RMathieu00}.  
For example, in the well-studied nearby low-mass star forming Taurus-Auriga 
region, the binary frequency of pre-main sequence stars is 
a factor of two higher than that of the main-sequence stars.
Other star-forming regions such as Ophiucus, 
Lupus, Chamaeleon, Scorpius-Centaurus, and Corona Australis 
show a similar excess of binary fraction.
Observations of binary stars in young star clusters such as 
Orion and Pleiades indicate that the binary frequency of the 
pre-main sequence stars is comparable to that of the main-sequence stars 
\citep{CProsser94,DPadgett97,JBouvier97}.
These observations suggest that binary formation is the
primary mode of star formation and that binaries should form
prior to the pre-main sequence phase, most likely through 
fragmentation during the gravitational collapse of the 
parent dense molecular cloud cores 
\citep{RMathieu94,PBodenheimer00,RWhite00}.
Recent detections of binaries and multiple systems 
at the earliest observable phases of star formation 
(e.g., Class 0 and Class I) support the fragmentation scenario
\citep[e.g.,][]{AWootten89,GFuller96, LLooney97, LLooney00, STerebey97}.
How the fragmentation occurs is a long-standing, unresolved problem. 

Many theoretical studies of star formation have concentrated on 
the formation of single stars \citep{FShu87}.  
In the standard picture of single star formation, a strongly
magnetized dense molecular cloud core is taken as the initial state 
of star formation.
The dense core gradually contracts as the magnetic support weakens 
in the high density part due to the ambipolar diffusion.
When the magnetic field becomes too weak, it is overwhelmed 
by self-gravity, and the cloud begins a dynamic contraction to form 
a star. The process of magnetic leakage during the cloud contraction 
has been extensively investigated both analytically 
\citep[e.g.,][]{LMestel56,TNakano72} and numerically
\citep[e.g.,][]{TNakano79, DBlack82, SLizano89, GCiolek93, SBasu94, 
ZLi01}. Specific models based on this scenario appear able to
explain the observations of well-studied dense cores B1 and 
L1544 
\citep[]{RCrutcher94,GCiolek00,ZLi02b}.
However, these studies have assumed axisymmetry of the model cloud, 
and are unable to address the issue of fragmentation. 
On the other hand, numerical hydrodynamic calculations of cloud
fragmentation have neglected the effects of magnetic fields
\citep[e.g.,][]{JTohline02}, which are potentially important 
or even critical. 

To quantify the role of magnetic fields on cloud fragmentation, we 
have made a start in investigating the non-axisymmetric evolution 
and fragmentation of initially magnetically subcritical clouds, 
taking into account the effect of magnetic diffusion. Our ultimate 
goal is to elucidate the fragmentation process in magnetically 
supported clouds, and extend the current standard scenario of 
low-mass star formation to binaries and multiple systems.

Magnetically subcritical clouds are stable to both dynamical collapse 
and fragmentation \citep[e.g.,][]{KTomisaka88}. Obviously, it is 
difficult to form binary or multiple stars by direct fragmentation 
of such clouds. Three-dimensional magnetohydrodynamic (MHD) simulations 
of magnetic cloud collapse have shown that a frozen-in magnetic field 
stifles cloud fragmentation \citep{EDorfi82,WBenz84,GPhillips86a,
GPhillips86b}.  In lightly ionized molecular clouds, however, ambipolar 
diffusion can initiate dynamical contraction in magnetically
supercritical cores by redistributing the magnetic flux in the cloud. 
The supercritical cores can be unstable to fragmentation if they 
have masses well above the Jeans limit \citep[e.g.,][]{LMestel56, DGalli01}. 

Recently, \citet{ABoss00, ABoss02} carried out three-dimensional collapse
calculations of initially magnetically supported clouds, taking into 
account several properties of magnetic fields approximately. He concluded 
that the cloud fragmentation is enhanced by magnetic fields because the 
magnetic tension allows more time for the perturbations to grow 
off center nonlinearly, which presents the formation of a single,
dominant object at the center. 
\citet{ZLi01} studied ambipolar diffusion-driven collapse of 
magnetized cloud cores with one-dimensional axisymmetric calculations
but treating the magnetic forces and ambipolar diffusion more 
accurately. He also found a similar beneficial effect of magnetic tension.
He classified the evolution of magnetically subcritical 
axisymmetric clouds into two types, 
depending mainly on the initial cloud mass and initial density 
distribution. When the initial cloud has a relatively low mass and/or a
centrally-condensed density distribution, it collapses to form a single
supercritical core ({\it core-forming cloud}).  
On the other hand, when the initial cloud is more massive (containing
several Jeans masses or more) and has a relatively flat density 
distribution, it collapses to form a ring after the central
region becomes magnetically supercritical ({\it ring-forming cloud}).
One-dimensional axisymmetric evolution of the core-forming clouds 
has also been extensively investigated by \citet{GCiolek93} and 
\citet{SBasu94}. \citet{ZLi02} examined the ring-forming clouds and
showed that they can fragment into multiple cores in the presence of 
non-axisymmetric density fluctuations during the isothermal collapse 
phase after the supercritical core formation. This type of ring 
fragmentation produces fragments of relatively large initial separations
($\sim 10^4$ AU). It may be responsible for the formation of small stellar 
groups such as those in the Serpens cloud core \citep{JWilliams00} and 
aggregates of pre-stellar cores such as those in the $\rho$ Oph B2 core 
\citep{FMotte98} and around IRAS 03256+3055 in Perseus (Young et al. 
2002). These studies illustrate the crucial role of magnetic diffusion 
in the fragmentation of magnetic clouds, with strong implications 
for binary formation.

In this paper, we focus on the non-axisymmetric evolution of the 
core-forming clouds, with an eye of the formation of binary stars. 
Initial results are presented in \citet{FNakamura02}, where we 
demonstrated that the core-forming clouds become unstable to 
a bar mode after the supercritical core formation [see also 
\citet{FNakamura97} for initially magnetically supercritical clouds]. 
Here, we extend the calculations to 
a larger parameter regime. First, we briefly describe our numerical 
method and model in \S \ref{sec:method}. In \S \ref{sec:results}, 
we examine the dependence of bar growth 
on the initial parameters such as 
magnetic strength, initial cloud mass, density profile, and rotation.
We show that the core-forming clouds become unstable to a bar mode
during the dynamic collapse phase over a wide range of initial model 
parameters. Subsequent fragmentation process is investigated in 
\S \ref{sec:fragmentation}. Based on the numerical results, we 
classify the fragmentation process into three types, which agree 
with the empirical classification of young binary and multiple 
systems by \citet{LLooney00} based on high-resolution millimeter 
observations. This general agreement and other implications of our 
results on binary formation are discussed in \S \ref{sec:discussion}.  

\section{Formulation of The Problem}
\label{sec:method}

A detailed description of the model cloud and the problem formulation are 
presented in \S 2 of \citet{ZLi02}.  Here, we give a brief summary. 
We consider an initially axisymmetric, magnetically subcritical model 
cloud that is in a mechanical equilibrium, with a disk-like mass 
distribution due to settling along ordered magnetic field lines. 
In such a subcritical cloud, force balance along the field lines
is achieved during the quasi-static phase of evolution and is well 
maintained even during the subsequent dynamic collapse 
\citep[e.g.,][]{RFiedler93, FNakamura95}.
Thus, we adopt the standard thin-disk approximation in which 
exact force balance along the vertical direction is assumed during
the entire evolution \citep[e.g.,][]{GCiolek93, SBasu94, FShu97, 
FNakamura97, RStehle01, ZLi01}.
The magnetohydrodynamical equations are integrated in the vertical
direction and the cloud evolution is followed in the ($x$-$y$)
disk plane of a Cartesian coordinate system $(x,y,z)$ with a
two-dimensional MHD code.  
The distribution of magnetic field is computed in 
three-dimensional space under the assumption that the magnetic field is
current-free outside the disk.

We assume an isothermal equation of state below a dimensionless surface 
density $\Sigma _{\rm cr}$, and denote the isothermal sound speed by
$c_s$. Beyond $\Sigma_{\rm cr}$, we change the equation of state
from isothermal to adiabatic, with an index of $5/3$,
to mimic the transition to optically thick regime of cloud evolution.
In this paper, we set $\Sigma_{\rm cr}=1.9\times 10^4 \Sigma _{\rm 0, ref}$, 
corresponding to the volume density $\rho_{\rm cr}\sim 10^{12}$ cm$^{-3}$
(see eq. [\ref{eq:sigma}] below for the definition of $\Sigma_{\rm 0,ref}$).

The initial conditions for star formation are not well determined
either observationally or theoretically. Following 
\citet{SBasu94} and \citet{ZLi01}, we prescribe a slowly-rotating,
 axisymmetric reference state for our model cloud,  
with the distributions of mass, magnetic field, and velocity
given by
\begin{eqnarray}
\Sigma _{\rm ref} (x,y) &=& \frac{\Sigma_{\rm 0,ref}}
{\left[1+(r/r_0)^n\right]^{4/n}} \ \ \ , \label{eq:sigma}\\
B_{z, \rm ref} (x,y) &=& B_{z, \infty} = {\rm const.}  \ \ \ , \\
V_\phi (x,y)&=&{ 4 \; \omega \; r \over r_0
	+\sqrt{r_0^2+r^2} } c_{ms}  ,
\end{eqnarray}
where $r= (x^2+y^2)^{1/2}$ is the distance from the origin, 
$c_{ms}=c_s (1+\Gamma_0^2)^{1/2}$ is essentially the magnetosonic 
speed, and the parameter $\omega$ measures the rotation rate.
The exponent $n$ controls the surface density profile and (indirectly) 
the amount of mass in the central plateau region 
where the distribution is more or less uniform.
The background field strength $B_{z, \infty}$
is characterized by the dimensionless
flux-to-mass ratio $\Gamma_0=B_{z, \infty}/(2\pi G^{1/2} \Sigma_{0, \rm
ref})$. 
The reference clouds are not in an equilibrium state and are
allowed to evolve into one with magnetic field frozen-in, before
ambipolar diffusion is turned on at time $t=0$. To save computational
time, we obtain the equilibrium state using the one-dimensional 
axisymmetric MHD code of \citet{ZLi01}, which gives results 
indistinguishable from those from the two-dimensional code assuming
axisymmetry. 

At $t=0$, we impose on top of the equilibrium surface 
density $\Sigma_0(x,y)$ an
$m=2$ perturbation of relative amplitude $A$, 
\begin{equation}
\Sigma (x,y)= \Sigma_0(x,y)\ [1 + A\; \cos(2 \phi)],
\end{equation}
where $\phi$ is the azimuthal angle measured from the $x$-axis. 
The subsequent, ambipolar diffusion-driven evolution of the perturbed
cloud is followed numerically with the two dimensional MHD code.
As a boundary condition, we fixed $\rho$ and $B_z$ at the cloud 
outer radius, taken to be twice the characteristic radius $r_0$.

The computations are carried out using non-dimensional
quantities. The units we adopted are $c_s$ for speed,
$\Sigma _{0,\rm ref}$ for surface density, $2\pi G\Sigma_{0,\rm ref}$
for gravitational
acceleration, and $B_{z,\infty}$ for magnetic field strength.
The units for length and time are, respectively, $L_0\equiv
c_s^2/(2\pi G\Sigma_{0,\rm ref})$ and $t_0\equiv c_s/(2\pi
G\Sigma_{0,\rm ref})$. They have typical values of 
\begin{eqnarray}
L_0 &=& 8.43 \times 10^{-2}\ {\rm pc}\left(\frac{c_s}
{0.33 \ {\rm \ km \ s^{-1}}}\right)^2
\left(\frac{\Sigma_{0,\rm ref}}{10^{-2} 
{\rm \ g\ cm^{-2}}}\right)^{-1}, \\
t_0 &=& 2.50\times 10^5 {\rm yr} \left(\frac{c_s}
{0.33 \ {\rm \ km \ s^{-1}}}\right)
\left(\frac{\Sigma_{0,\rm ref}}{10^{-2} {\rm \ g\ cm^{-2}}}\right)^{-1}, 
\end{eqnarray}
where $c_s=0.33$~km~s$^{-1}$ corresponds to an effective temperature of
$T_{\rm eff}=30$~K. 

The initial state of our model cloud is completely specified by four 
parameters: the characteristic radius $r_0$, exponent $n$ in the 
reference surface density distribution, dimensionless flux-to-mass 
ratio $\Gamma_0$, and rotation parameter $\omega$. 
We adopt a relative small rotation parameter of $\omega \lesssim 0.2-0.3$,
consistent with the slow rates of rotation observed in molecular cloud 
cores \citep{AGoodman93}. The slow rotation is thought to be due to 
magnetic braking by an ambient medium \citep{TNakano72,SBasu94}, which 
is not treated in our paper. We comment briefly on the potential effects 
of magnetic braking in \S~\ref{types}.

Besides the four parameters characterizing the initial cloud, a parameter 
$A$ is needed to specify the fractional amplitude of the 
$m=2$ bar mode perturbation applied after the reference cloud settles
into the equilibrium configuration. We have carried out a systematic 
parameter survey and selected the models listed in 
Table~\ref{tab:1} to illustrate the salient 
features of magnetic cloud evolution. The numerical results are described 
in the next two sections (\S \ref{sec:results} and 
\S \ref{sec:fragmentation}).

\section{Bar Growth in Magnetically Subcritical Clouds}
\label{sec:results}

As mentioned earlier, \citet{ZLi01} classified the evolution of magnetically 
subcritical axisymmetric clouds into two types, 
depending mainly on the initial cloud mass and initial density 
distribution: (a) {\it core-forming cloud}, 
which collapses to form a single supercritical core and 
(b) {\it ring-forming cloud} in which a supercritical ring is formed
after the central region becomes supercritical owing to ambipolar diffusion.
The ring fragments into multiple supercritical cores 
in the presence of density fluctuations 
during the isothermal collapse phase \citep{ZLi02}.
In this paper, we focus on the core-forming clouds and 
show that these clouds become unstable to bar formation 
after the supercritical core formation over 
a wide range of initial model parameters.
The subsequent evolution and fragmentation of the bar are 
discussed in \S \ref{sec:fragmentation}.

\subsection{Non-Rotating Cloud}
\label{subsec:norotation}

\begin{figure}
\epsscale{1.0}
\plotone{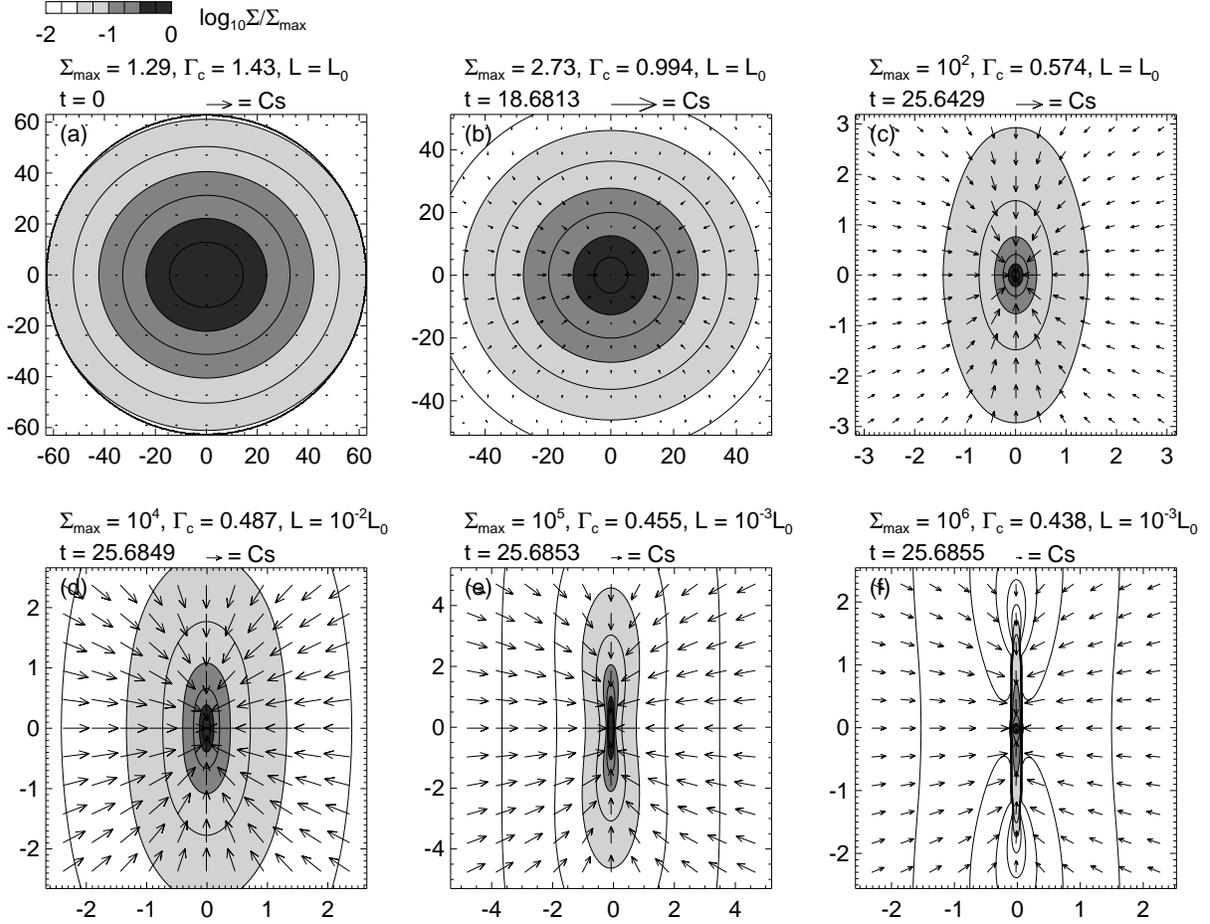}
\caption{ Snapshots of the 
surface density distribution and velocity field for the
non-rotating cloud perturbed by an $m=2$ mode of fractional amplitude 
$A=0.05$ at six dimensionless times: (a) t=0, (b) 18.6813, (c)25.6429, (d)
 25.6849, (e) 25.6853, and (f) 25.6855.
The model cloud has a set of ``standard'' parameters 
of $(r_0, n, \omega, \Gamma _0)=(10\pi, 2, 0, 1.5)$.
Panel (a) shows the initial state.  
In panels (b) through (f), only the central regions are shown.
The contour curves in each panel are for the surface density normalized by 
$\Sigma_{\rm max}$, whose value is given above each panel. Also
given are the flux-to-mass ratio ($\Gamma_c$) at the
density peak, and length unit for each panel.
The arrows are velocity
 vectors normalized by the effective isothermal sound speed $c_s$ 
(without magnetic contribution), whose magnitude is indicated 
above each panel. Note that in panel (a) the added perturbation 
lowers the flux-to-mass ratio $\Gamma=1.43$ from the reference value 
$\Gamma_0=1.5$ by $5\%$, corresponding to the fractional amplitude 
of the perturbation. 
For dimensional units, see \S~\ref{sec:method}.
\label{fig:1}}
\end{figure}

To begin with, we consider a non-rotating ($\omega=0$) cloud with the
reference state specified by the set of parameters $r_0=10\pi$, $n=2$,
and $\Gamma_0=1.5$. The cloud is therefore magnetically subcritical,
with the field strength $50\%$ above the critical value at the center.
The cloud so specified produces a dense supercritical core 
in the absence of any non-axisymmetric perturbations.
The reference cloud is allowed to evolve into an equilibrium
configuration, with the magnetic field frozen-in. 
After the equilibrium is reached, we reset the time to $t=0$, 
turn on ambipolar diffusion, and add an $m=2$ perturbation to the 
surface density distribution, 
with a fractional amplitude of $A=0.05$. 
The evolution of such a non-axisymmetrically perturbed cloud is 
followed numerically to progressively higher densities 
and smaller scales. The results are 
shown in Fig.~\ref{fig:1} and described as follows.

The evolution of the model is qualitatively similar to that of the model
shown in Fig. 1 of \citet{FNakamura02}, where a different
reference density distribution (less centrally-condensed than the model
shown in this subsection) is adopted. The evolution is followed to
a higher density here. 
As is usual with ambipolar diffusion-driven evolution, the cloud spends
most of its time in the subcritical phase when the dimensionless
minimum flux-to-mass ratio (at the density peak) is greater than unity,
$\Gamma _c > 1$ (see panels [a] and [b] of Fig.~\ref{fig:1}).
During this period, gas motions are very subsonic and the central part
of the cloud oscillates with a small amplitude. 
The oscillation indicates that the cloud is stable to the
$m=2$ perturbation during the subcritical phase, in agreement with
linear analysis by Nakano (1988) and others. Indeed, we have verified
numerically that for a frozen-in field the subcritical cloud remains
stable to this and other higher (e.g., $m=3,4, \dots$) modes.  
(We followed the evolution of the frozen-in model 
for several periods of oscillation and confirmed that the cloud 
oscillates with almost the same amplitude as the initial.)
Once the minimum flux-to-mass ratio has dropped below the critical
value, the contraction becomes dynamic and the infall speed 
eventually exceeds the sound speed at the center (see panels [c] and 
[d]). 
By the time shown in panel (c), the $m=2$ mode has grown significantly, 
resulting in a bar-like core at the center with an aspect ratio of roughly 2.
The bar growth remains relatively modest (panel [d]) 
until the very end of the 
runaway 
collapse, when the growth rate increases dramatically  (panel [e]).  
Beyond a dimensionless critical surface density 
of $1.9\times 10^4$, corresponding to a volume density of $n_H\sim10^{12}$
cm$^{-3}$, we change the equation of state from isothermal to adiabatic,
with an index of 5/3, to mimic the transition to optically thick regime
of cloud evolution.  
When the maximum surface density reaches $\sim 10^5$, the contraction
along the minor axis of the bar decelerates significantly 
and an accretion shock is formed at the surface of the bar.  
This shock-bound bar is analogous to the ``first core'' of  
spherical calculations, and will be referred to as the ``first bar''
below.  Immediately after the shock formation, 
the aspect ratio of the first bar continues to increase. It  
reaches a maximum of about 10 by the time shown in panel (e).
Thereafter, the infall along the major axis becomes significant, 
forming a round core at the center.  
The aspect ratio of the central round core approaches unity 
as the collapse proceeds.   Since the round core resembles 
the first core of the spherical calculations, we call it the
``truly first core''. As will be shown in \S \ref{sec:fragmentation}, 
this round core evolves into a rapidly-rotating disk in the presence 
of an initial, slow cloud rotation. 

\begin{figure}
\epsscale{0.50}
\plotone{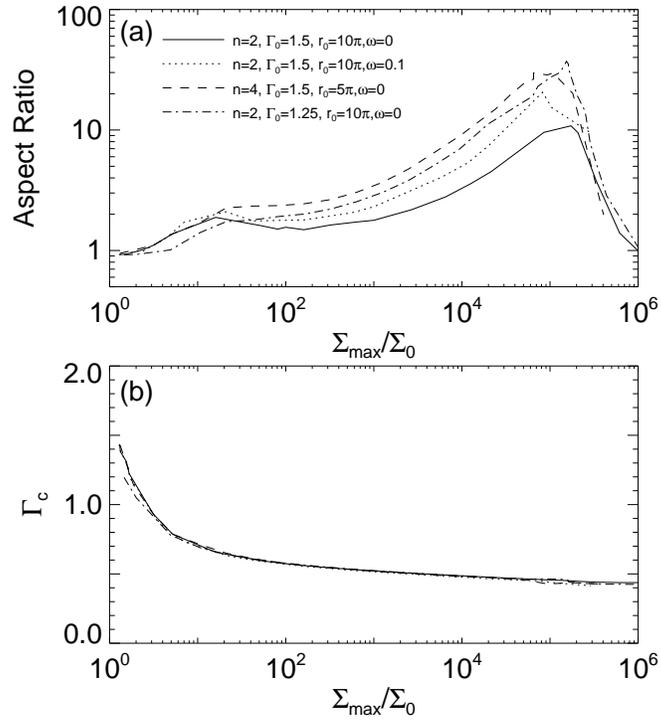}
\caption{ Evolution of (a) the aspect ratio of the highest density region
with surface density above $10^{-1/2} \Sigma_{\rm max}$, and (b) the 
flux-to-mass ratio $\Gamma_c$ at the surface density maximum.
Solid curves are for the standard model with $n=2$, $\Gamma _0 = 1.5$, 
$r_0=5\pi$, and $\omega = 0$.
\label{fig:2}}
\end{figure}

To quantify the geometrical change of the supercritical core, we plot in 
Figure \ref{fig:2} the aspect ratio of the darkest region in  Figure
\ref{fig:1} (with $\Sigma > 10^{-1/2}\Sigma_{\rm max}$)
as a function of the maximum surface density, $\Sigma_{\rm max}$, 
together with the flux-to-mass-ratio $\Gamma$ at the center.
The aspect ratio is defined as $R = a_y/a_x$
where $a_x$ and $a_y$ are, respectively, the effective scale heights 
 along the $x$ and $y$ axes, and evaluated as 
\begin{equation}
a_x = \left(M^{-1}\int_{\Sigma \ge 10^{-1/2} 
\Sigma _{\rm max}} \Sigma x^2 dx dy
\right)^{1/2},
\end{equation}
and 
\begin{equation}
a_y = \left(M^{-1}\int_{\Sigma \ge 10^{-1/2} \Sigma _{\rm max}} 
\Sigma y^2 dx dy
\right)^{1/2},
\end{equation}
where $M$ is the total mass in the region with $\Sigma \ge 10^{-1/2} 
\Sigma _{\rm max}$.
During the subcritical phase, the aspect ratio of the bar does not grow
in time; it oscillates around unity, signifying that the subcritical
cloud is stable to the $m=2$ mode.  
As the central region becomes more and more supercritical, 
the bar mode begins to grow significantly.
During this supercritical period, the aspect ratio of the bar 
stays more or less frozen at $R\sim 1.5$ before the central density
reaches $\Sigma \sim 10^3$.
This freezing of aspect ratio is related to the dynamical state
of the disk.  
As shown in \S \ref{subsec:why},
the mass of the central flat region is merely $\sim$ 1.5 times 
the effective Jeans mass (as measured by the central surface
density, including the magnetic contribution).
Such a relatively low-mass disk is only marginally unstable to the
non-axisymmetric perturbation.  In other words, the time scale of  
bar growth is comparable to that of dynamic contraction.
The flux-to-mass ratio does not change much either during this period,
as a result of the rapid collapse which prevents the magnetic flux 
from leaking out significantly. The bar growth speeds up toward the very 
end of the 
supercritical collapse when the mass of the central flat region 
exceeds about twice the effective Jeans mass. It continues after 
the first bar formation.
When the central surface density reaches $10^5$, the infall motion
along the major axis of the bar becomes significant and 
the aspect ratio begins to drop steeply to unity, 
forming the truly first core.

\begin{figure}
\epsscale{1.0}
\plotone{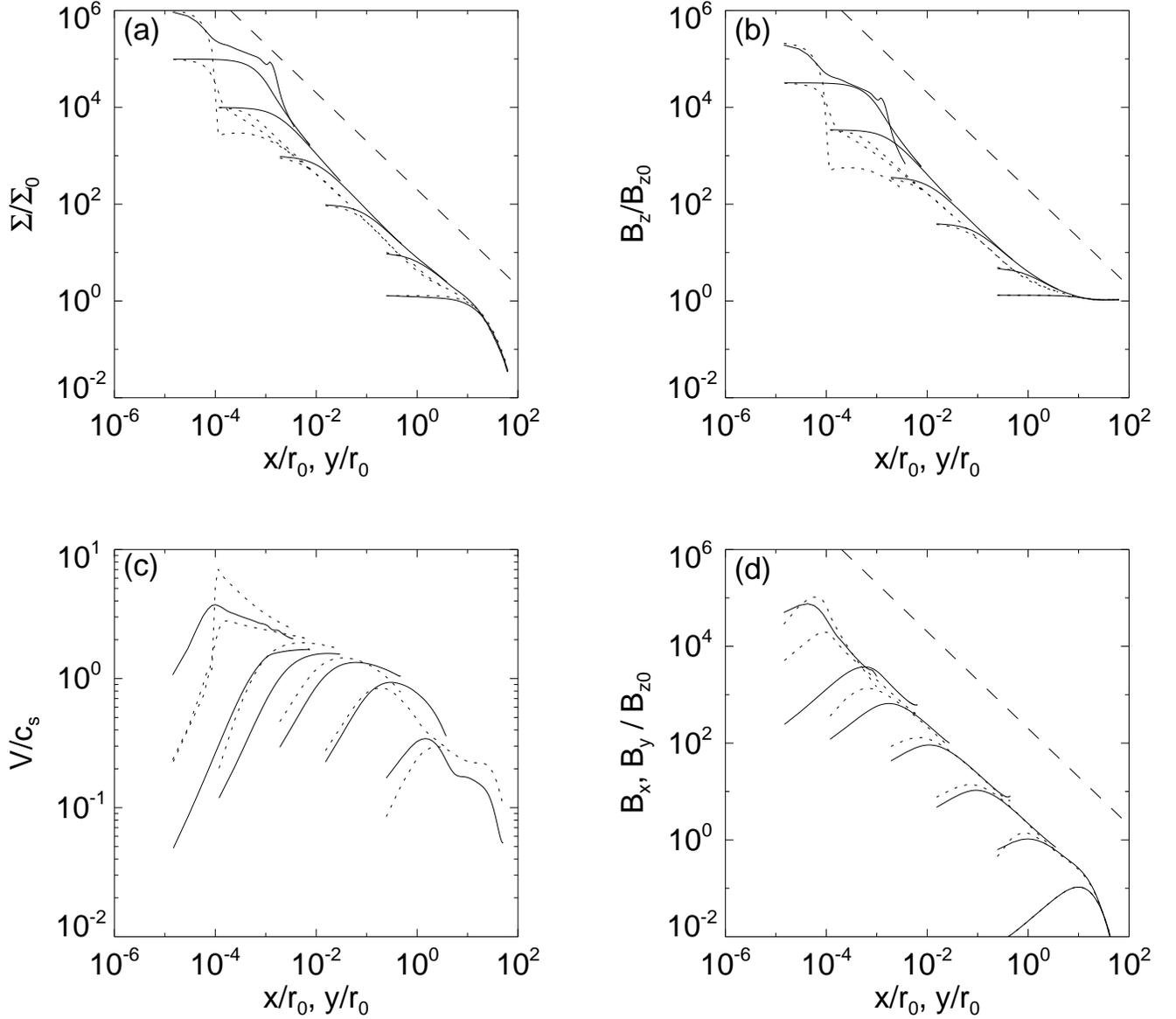}
\caption{Distributions of surface density $\Sigma$, infall speed $V$, and 
the vertical magnetic field strength $B_z$, and the two components of 
magnetic field in the disk plane ($B_x$ and $B_y$) along the major (solid 
lines) and minor (dotted lines) axes of the standard model shown in 
Fig.~\ref{fig:1}, at seven different times. Note that before the breakdown
of isothermal equation of state the surface density and the strength of 
each magnetic field component approach a power-law distribution of 
$\propto d^{-1}$ (plotted as a dashed line in panels [a], [b], and [d],  
with $d$ denoting the distance from the center), and the infall speed 
is more or less constant. This self-similar behavior of
isothermal collapse is explored further in \S \ref{subsec:why}. 
In panels (b) and (d), the solid and dotted lines coincide with each
 other at the initial state because of the initial axisymmetric magnetic 
configuration.  In panel (c), the initial infall speed is equal to zero
 owing to the static initial condition and accordingly not depicted.
\label{fig:3}}
\end{figure}

To study the structure of the bar, we show in Figure \ref{fig:3}
the distributions of the surface density, magnetic field, and infall
velocity along the axes as a function of the distance 
from the density maximum, at seven different times.
The solid and dashed curves indicate the distributions of the 
physical quantities along the major ($y$) and minor ($x$) axes, 
respectively.  
During the early phase of supercritical collapse when the aspect ratio
of the bar is frozen at $R\sim 1.5$, the surface density distributions 
along both axes qualitatively resemble the self-similar distribution of 
the axisymmetric collapse; they are flat near the density maximum 
and can be approximated 
by a power-law of $\Sigma \propto d^{-1}$ at the magnetically 
supercritical envelope, where $d$ is the distance from the density peak. 
The $\Sigma \propto d^{-1}$ distribution corresponds to 
a volume density distribution of $\rho \propto d^{-2}$ because of the
force balance in the vertical direction (which yields $\rho\propto 
\Sigma ^2$).
While the central density increases rapidly, the surface density
distributions change little in the envelope.  The central flat region
shrinks in size in accordance with the increase in the central density.
The distributions of the vertical field $B_z$ are qualitatively 
similar to those of
$\Sigma$ in the magnetically supercritical region and 
almost proportional to the local surface density ($B_z \propto \Sigma$).
The $x$- and $y$-components of the magnetic field (in the disk plane) 
increase approximately linearly with distance $d$ in the central flat 
region, peaking near the edge of the flat region, before 
falling off as $B_x \propto d^{-1}$ and $B_y \propto d^{-1}$.
In the magnetically supercritical envelope, $B_x$ and $B_y$ are 
comparable to $B_z$, indicating that the field lines near the disk plane 
incline with an angle of $\sim$ 45 degree with respect to the disk plane.
The angle approaches $\sim 90$ degree in the outer, magnetically
subcritical envelope. 
The infall speeds are proportional to the distance $d$ 
near the center, becoming more or less flat in the supercritical
part of the envelope, before falling to subsonic values in 
the outer subcritical envelope.  
At $\Sigma_{\rm max} = 10^4$, the infall speeds in the supercritical
envelope reach $\sim 2c_s$.

As the collapse proceeds and the bar aspect ratio becomes large, 
the distributions of physical quantities gradually deviate from the
self-similar distributions.  For example, the surface density
and vertical magnetic field ($B_z$) profiles along the major axis 
become steeper as the bar elongates rapidly. 
When the central surface density reaches $\sim 2\times 10^4$, 
a shock is formed at $x\sim 10^{-4} r_0$.
Thereafter, the infall speed rises in the envelope of the bar.
The infall speed along the minor axis follows $v_x \propto d^{-1/2}$
just outside the first bar, 
indicating a free-fall collapse towards the bar.

Interestingly, after the bar becomes opaque to dust emission, 
a local density peak is formed at each end of the first bar.
This density peak formation is due to the ``edge'' effect,
which was first calculated by \citet{RLarson72} and then extensively
investigated by \citet{PBastien83} and \citet{IBonnell91}.
In this model, the density peaks do not grow nonlinearly;  
they tend to disappear in the subsequent collapse because 
the central truly first core grows more rapidly in mass
owing to the supersonic infall along the major axis, 
which dominates the gravitational potential in the high density region.
In \S \ref{sec:fragmentation}, we show that for more massive clouds 
in which the aspect ratios of the supercritical cores are larger, 
the density peaks {\it can} grow in time because the infall speeds 
along the major axis remain subsonic. 

The model discussed in this subsection serves as a standard against 
which other models will be compared. 

\subsection{Dependence of Model Parameters: $\Gamma _0$, $r_0$, and $n$}
\label{subsec:gamma}

The properties of the bar growth depend on the initial model parameters.
Here, we investigate the effects of the initial flux-to-mass ratio
$\Gamma _0$, the characteristic radius $r_0$, and the density profile
parameter $n$ for non-rotating clouds.
The effect of rotation is examined in separate subsections 
\S \ref{subsec:rotation} and \S \ref{subsec:separate core}. 
Exact values of $\Gamma_0$ are difficult to
determine observationally; they probably lie within a factor of two 
of unity \citep{RCrutcher99}, after correcting for likely projection 
effects \citep{FShu99}.  
Thus, we investigated the dependence on $\Gamma _0$ in the range 
of $1\lesssim \Gamma _0 \lesssim 3$.
The characteristic radius $r_0$ is to be 
compared with $2\pi$, the (dimensionless) critical 
wavelength for the (thermal) Jeans instability in a disk of uniform 
mass distribution \citep{RLarson85}. The parameter $n$ controls the 
profile of the cloud surface density distribution; those with smaller 
values of $n$ are more centrally peaked.

\begin{figure}
%\epsscale{0.6}
\plotone{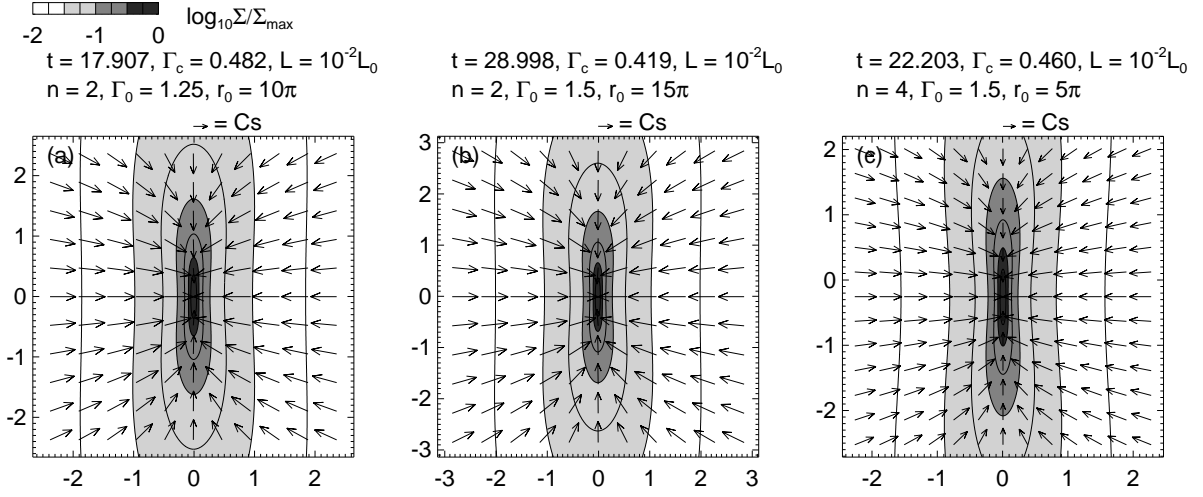} 
\caption{Snapshots of the surface density distribution and velocity
field at the time when $\Sigma _{\rm max} = 10^4$ for 
the nonrotating models with different model parameters.
Panel (a): the model with a weaker magnetic field,
$(r_0, n, \omega, \Gamma _0, A)=(10\pi, 2, 0, \underline{1.25}, 0.05)$,
showing that decreasing field strength promotes bar growth. 
Panel (b): the model with a larger $r_0$, 
$(r_0, n, \omega, \Gamma _0, A)=(\underline{15\pi}, 2, 0, 1.5, 0.05)$,
showing the positive effect of the initial cloud mass on the bar growth.
Panel (c): the model with a larger $n$ and a smaller $r_0$, 
$(r_0, n, \omega, \Gamma _0, A)=(\underline{5\pi}, \underline{4}, 
0, 1.5, 0.05)$, showing the positive effect on the bar growth 
of a flat density distribution.
These panels should be compared with panel (d) of Fig. \ref{fig:1}.
The contours, arrows and notations have the same meaning as in 
Fig.~\ref{fig:1}.
The parameters having different values from those of Fig. \ref{fig:1}
are underlined.
\label{fig:4}}
\end{figure}

To examine the effect of the initial flux-to-mass ratio $\Gamma_0$, 
we first hold the cloud radius and profile parameter 
fixed at the standard values $r_0=10\pi$ and $n=2$, 
and reduce the flux-to-mass ratio $\Gamma _0$.
Representative results are shown 
in panel (a) of Fig. \ref{fig:4}, 
at a time when the maximum surface density 
$\Sigma_{\rm max}=10^4$.
Panel (a) is the snapshot of a cloud 
with $\Gamma_0=1.25$ (a weaker initial magnetic field),
perturbed by an $m=2$ mode of the 
standard fractional amplitude $A=0.05$. 
Comparison with panel (d) of Fig. \ref{fig:1} where the
standard case at the same $\Sigma_{\rm max}$ is plotted 
shows that a weaker magnetic field promotes the bar growth.  
This slower bar growth in the stronger field case 
is quantitatively shown in panel (a) of Figure \ref{fig:2}
(compare the solid curve with the dotted-dashed curve). The
reason for this behavior appears to be that, for the cloud 
with a stronger initial magnetic field, the magnetic support 
remains more important even in the supercritical phase. 
In other words, the averaged flux-to-mass ratio in the darkest region
($\Sigma \ge 10^{-1/2}\Sigma _{\rm max}$) is larger for the stronger
field model. The stronger field tends to retard the deformation 
of the supercritical core. We have confirmed this negative effect of the
magnetic field on bar growth by following the evolution of the clouds 
with $\Gamma_0 = 2$ and 3.\footnote{We note that the magnetic field 
has an opposite, beneficial effect on the fragmentation of ring-forming 
clouds. More strongly magnetized clouds contract more slowly, allowing 
more time for non-axisymmetric perturbations to grow off-center. 
As a result, multiple fragmentation is achieved more readily, 
at a lower surface density.  See \citet{ZLi02} for details.}

We next consider the effect of the characteristic radius $r_0$ which 
controls the total cloud mass.  
Increasing the radius $r_0$ is expected to have a 
positive effect on bar growth.   This is indeed the case, 
as illustrated in panel (b) of Fig. \ref{fig:4}, 
where the snapshot of a cloud with $r_0=15\pi$, 
1.5 times the standard value, is shown at $\Sigma _{\rm max} = 10^4$.   
The aspect ratio of the bar reaches $\sim 7$ at the time when
the surface density reaches $10^4$.

The profile parameter $n$ has an effect similar to the parameter $r_0$.
For instance, a reduction in $n$ would decrease the number of 
Jeans masses contained in the central plateau region.
We showed in a previous paper \citep{ZLi02} that for the ring-forming 
case, the clouds with more peaked density profiles are 
less prone to fragmentation.
This negative effect of the centrally condensed density distribution is
also observed for non-magnetized clouds \citep{ABoss93}. 
We find the same trend for the core-forming clouds as well. 
The aspect ratio of the bar increases more rapidly for larger $n$.
This is illustrated in panel (c) of Fig. \ref{fig:4}, 
where $n=4$, twice the standard value, is chosen. The bar 
is more elongated than the standard case, even though  
the characteristic radius $r_0=5\pi$ is only half of the 
standard value.

\subsection{Slowly-Rotating Cloud}
\label{subsec:rotation}

\begin{figure}
%\epsscale{0.6}
\plotone{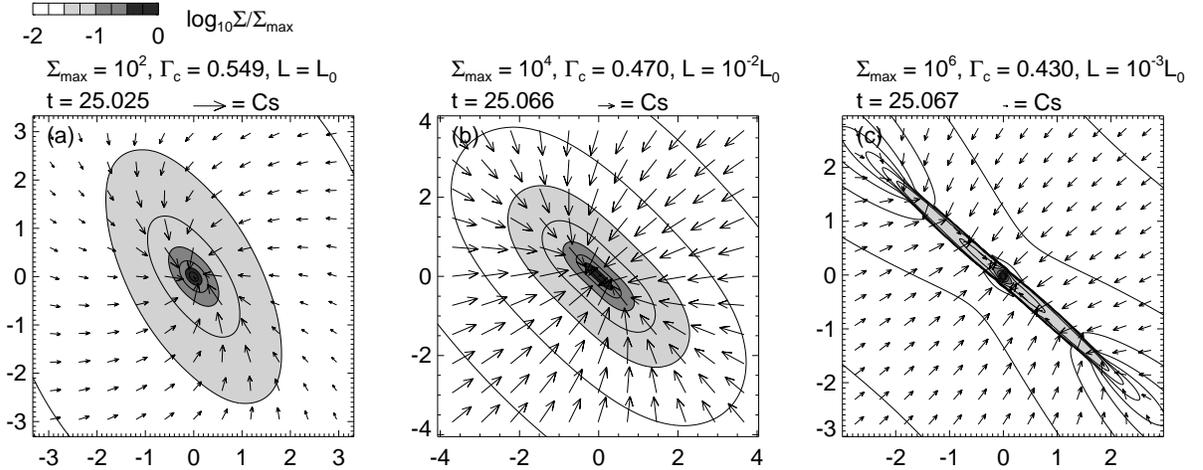} 
\caption{Snapshots of the surface density distribution and velocity
field of the slowly-rotating model with $(r_0, n, \omega, \Gamma _0, A)=
(10\pi, 2, \underline{0.1}, 1.5, 0.05)$ at three different stages: 
(a) $t = 25.025$, (b) $t = 25.066$, and (c) $t = 25.067$.
The contours, arrows and notations have the same meaning as in 
Fig.~\ref{fig:1}.
The parameters having different values from those of Fig. \ref{fig:1}
are underlined. Comparison with panels (c) through (e) of Fig. \ref{fig:1} 
points to a positive effect of the slow rotation on the bar growth.
\label{fig:5}}
\end{figure}

To gauge the effect of slow rotation, we repeat the evolution of a
cloud with all parameters identical to those of the standard model 
in \S \ref{subsec:norotation} except the rotation parameter $\omega$, 
which is now set to $0.1$ instead of zero. 
Snapshots of the surface density distribution and velocity
field at three selected times are shown in Fig.~\ref{fig:5}. 
The general trend of
the cloud evolution is similar to that in the non-rotating case;
in the subcritical phase, the cloud is stable to non-axisymmetric
perturbations and the $m=2$ mode begins to grow in time only after  
the central region becomes magnetically supercritical.
Comparison between Fig.~\ref{fig:1}d and Fig.~\ref{fig:5}b indicates 
a beneficial effect of the slow rotation on the bar growth. 
More quantitatively, the aspect ratio of the bar is 
always larger for the rotating cloud at the time 
when the maximum density reaches the same value, before the 
bar becomes opaque to dust emission. This effect derives 
from the fact that rotation retards the radial infall, which 
allows more time for perturbations to grow.

In \S \ref{subsec:separate core} below, we will show that the slow-down 
effect due to rotation plays a dramatic role in the cloud fragmentation
for more rapidly-rotating, less centrally-condensed clouds,  
which can fragment into two cores during the isothermal collapse phase 
soon after the supercritical core formation.

\subsection{Dynamical State of Supercritical Cores: Axisymmetric Collapse}
\label{subsec:why}

\begin{figure}
%\epsscale{0.6}
\plottwo{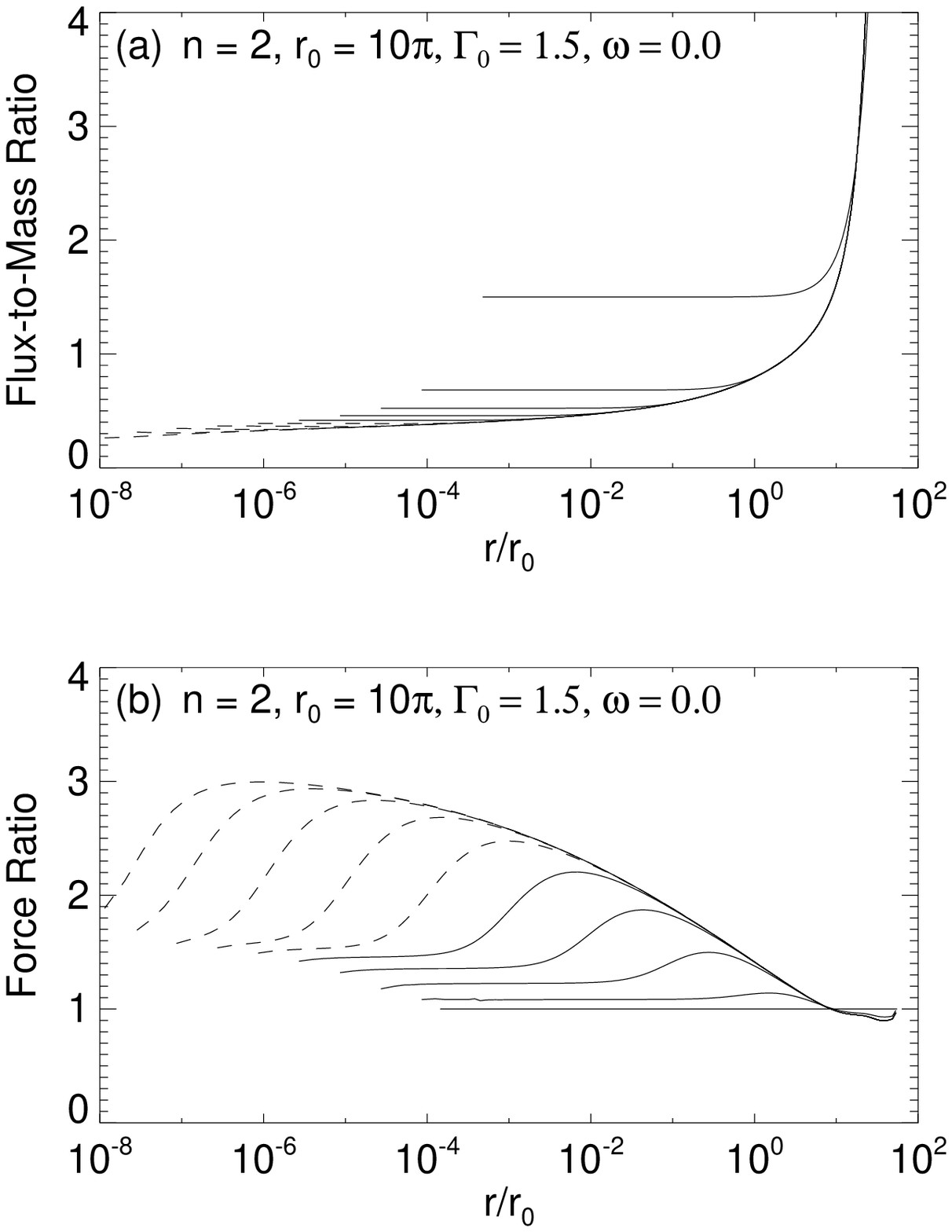}{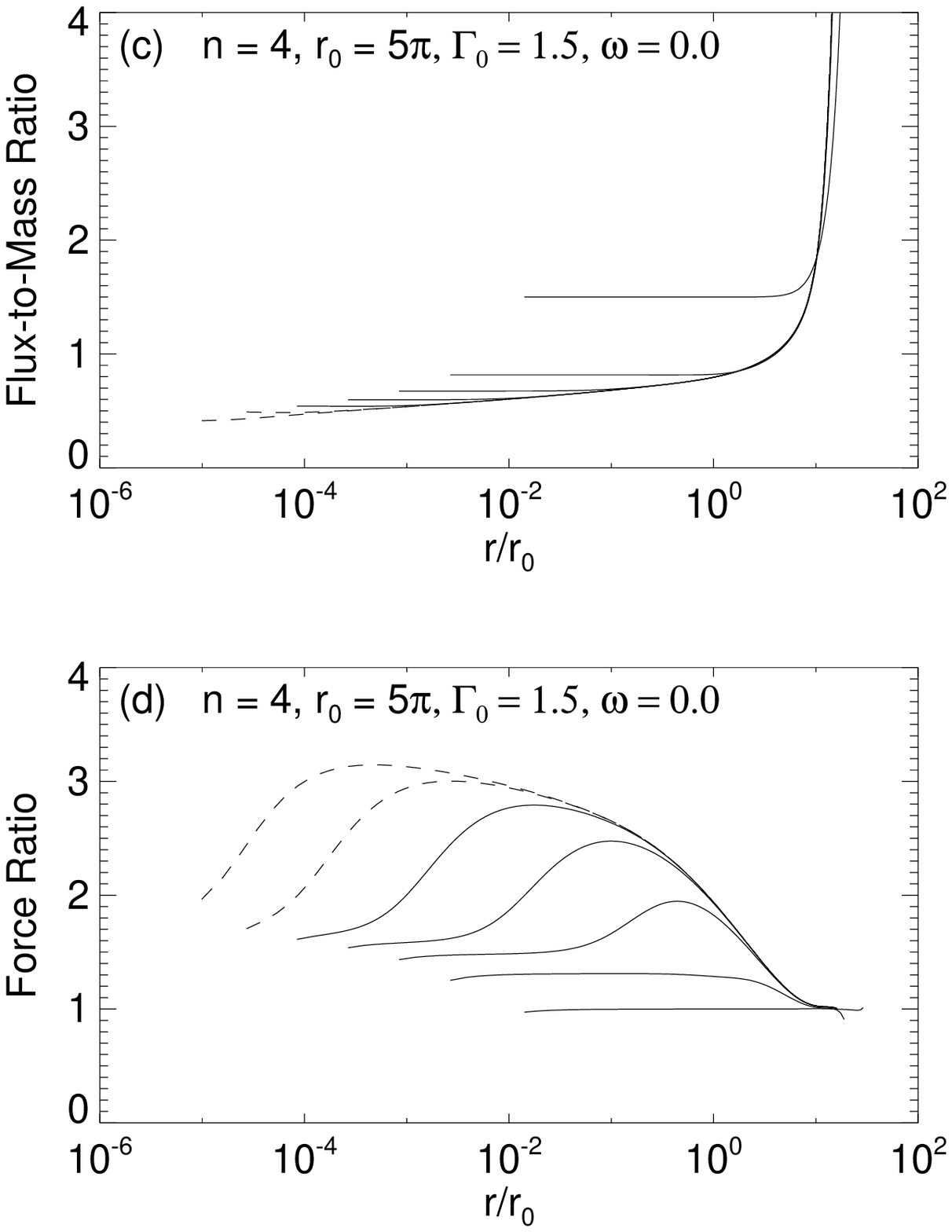} 
\caption{Axisymmetric evolution of two clouds without perturbations. Plotted
are (a) 
the flux-to-mass ratio and (b) the force ratio for a centrally-condensed
model with the standard set of cloud parameters $(r_0, n, \omega, \Gamma _0)
= (10\pi, 2, 0, 1.5)$, and (c) the flux-to-mass ratio and (d) 
the force ratio for a less centrally-condensed model with $(r_0, n, \omega, 
\Gamma _0)= (\underline{5\pi}, \underline{4}, 0, 1.5)$,  
at times when $\Sigma_c=1.2 (t=0), 10, 10^2, 10^3, \cdots$. 
The force ratio is defined as the ratio of the effective 
gravity ($F_{\rm g, eff}$) to the
 effective pressure force ($F_{\rm P, eff}$), where
$F_{\rm g, eff}=F_{\rm grav} + F_{\rm ten}$ and 
$F_{\rm P, eff}=F_{\rm gas} + F_{\rm B}$.
In these one-dimensional calculations, we do not stiffen 
the isothermal equation of state even when the surface density
 exceeds the critical value of $\Sigma _{\rm cr} = 1.9\times 10^4$.
In all panels, we show the distributions at the times when 
$\Sigma_c> \Sigma_{\rm cr}$ separately with dashed lines. 
\label{fig:1d}}
\end{figure}

The bar growth is closely related to the dynamical state of the
supercritical core.  In this subsection, to clarify the physics
of the bar growth, we follow the evolution of unperturbed axisymmetric 
clouds with the one-dimensional code and 
measure the dynamical forces in the supercritical cores.
For simplicity, we do not change the equation of state for the
one-dimensional calculations even when the surface density exceeds the
critical value of $1.9\times 10^4$. Previous axisymmetric calculations 
of ambipolar diffusion-driven cloud evolution have shown 
that the collapse of supercritical cores proceeds self-similarly: 
the distributions of the surface density and magnetic field are almost
constant in the central flat region and decrease with radius as 
$\Sigma \propto r^{-1}$ and $B_z \propto r^{-1}$ \citep{GCiolek93, 
SBasu94}. In the high density region of the supercritical core, 
the decrease of the flux-to-mass ratio levels off (at $\Gamma \sim 
0.4$) due to the rapid dynamic collapse which prevents significant 
magnetic flux leakage. This is shown in panels (a) and (c) of 
Figure \ref{fig:1d}, where the flux-to-mass ratio normalized by the
critical value is depicted as a function of radius for two 
representative models. 
\citet{FShu97} and \citet{ZLi97} called such a cloud with a constant
flux-to-mass ratio ``isopedic''.
In the isopedic cloud, the gravity ($F_{\rm grav} = \Sigma g_r$) 
and gas pressure ($F_{\rm gas}=c_s^2 d\Sigma/dr$) are proportional to 
the magnetic tension ($F_{\rm ten}= B_r B_z/ 2\pi$) and pressure
($F_{\rm B}=Hd(B_z^2/4\pi)/dr$), respectively \citep{FShu97, ZLi97,
FNakamura97}, where $H$ is the half thickness of the disk
(see eq. [5] of \citet{ZLi02} for the definition of $H$). 
In fact, the supercritical cores have such a characteristic.

Figure \ref{fig:1d} shows the evolution of the ratio of the effective 
gravity to the effective pressure at six different times for two 
models.  Here, the effective pressure and
gravity are defined as $F_{\rm gas} + F_{\rm B}$ and $F_{\rm grav} + 
F_{\rm ten}$, respectively.
At the initial state ($t=0$), this ratio is equal to unity, indicating
a force balance in the initial cloud.
As the collapse proceeds, this ratio increases due to ambipolar
diffusion, which reduces the magnetic support in the central high density
region. When the central surface density is in the range of $10^3-10^4$, 
the effective gravity is larger than the effective pressure, but only 
by a factor of about 1.5 $-$ 2. This corresponds to a mass of the central 
flat region only about 1.5 $-$ 2 times the effective Jeans mass.
Such a relatively low-mass core is only marginally unstable to 
non-axisymmetric perturbations, which is probably the reason why 
the aspect ratio of the bar remains almost frozen at a modest 
value of $1.5-2$ in the presence of an $m=2$ perturbation.  
When the force ratio exceeds $\sim 2$, the bar growth is accelerated 
remarkably by the Lin-Mestel-Shu type gravitational 
instability \citep{CLin65}, 
as seen in Fig. \ref{fig:2}. 
For both clouds, the force ratios approach a higher value 
of $3-4$ as the collapse proceeds, 
although a convergence hasn't been reached by the time 
the central surface density reaches the critical value of 
$1.9\times 10^4$.  
(The convergence toward a self-similar solution is slower for more
centrally-condensed clouds.)
This behavior suggests that the isothermal evolution 
tends toward a self-similar solution.

A self-similar solution reproducing the pre-protostellar collapse 
of an isopedic magnetic disk is derived approximately by \citet{FNakamura97}, 
who considered a frozen-in magnetic field [see also
\citet{KSaigo98} for the nonmagnetized case].
\citet{SBasu97} constructed a self-similar collapse model including the
effect of ambipolar diffusion [see \citet{ZLi98} for the self-similar 
collapse after the protostellar core formation]. The behavior of his 
solution is in good agreement with that of \citet{FNakamura97}
in the late, self-similar phase of dynamic collapse, when 
the effective gravity is about $3-4$ times the effective pressure force
and the asymptotic velocity is about twice the effective sound speed.
\citet{FNakamura97} investigated the dynamical stability 
of the self-similar solution to non-axisymmetric
perturbations and found that the self-similar solution is unstable 
to the $m=2$ mode and stable to the higher modes.
This characteristic is consistent with the numerical results presented
in the previous subsections. 
Thus, we conclude that the tendency for the supercritical collapse 
to approach the gravity-dominated self-similar solution is 
responsible for the bar formation during the dynamic collapse. 

\section{Fragmentation of Supercritical Cores}
\label{sec:fragmentation}

In \S \ref{sec:results}, 
we demonstrated that initially magnetically subcritical clouds tend 
to be unstable to an $m=2$ mode, forming bar-like cores. In this 
section, we investigate the subsequent evolution of the bar mode and 
show that the elongated supercritical cores can break up into discrete 
fragments. We classify the fragmentation process into three types, 
depending on the initial model parameters:
(1) {\it separate core formation}, in which the $m=2$ mode breaks up 
into two separate cores in the isothermal collapse phase, 
(2) {\it bar fragmentation}, in which the first bar breaks up  
into multiple fragments that move mainly along the major axis of the 
bar, with frequent merging among the fragments, and 
(3) {\it disk fragmentation}, in which the central region evolves 
into a rapidly-rotating disk where another bar mode is excited, 
fragmenting into blobs. In the following, we select three models 
to illustrate the salient features of these cases.

\subsection{Separate Core Formation: Effect of Faster Rotation}
\label{subsec:separate core}

In \S \ref{subsec:rotation}, we showed that rotation slows down the 
dynamic collapse, allowing more time for perturbations to grow.
Here, we demonstrate that a relatively fast rotation can affect 
the outcome of cloud evolution dramatically, particularly for 
$ n\gtrsim 4$ clouds whose initial parameter range is intermediate 
between those for core-forming and ring-forming clouds in the 
axisymmetric calculations \citep{ZLi01}.
Panels (b) and (c) of Fig. \ref{fig:6} show the snapshots 
of the surface density distribution and velocity 
field at two selected times for the model with $n=4$, 
$\omega =0.2$, $r_0=5\pi$, and $\Gamma_0=1.5$.
This cloud is categorized as a core-forming cloud from 
the axisymmetric one-dimensional calculation.
A cloud larger in radius by only a few tens of percent 
would collapse to form a ring instead of a single core.
For comparison, the snapshot of a more slowly rotating cloud
with $\omega = 0.1$ is shown in panel (a) at the time when 
$\Sigma _{\rm max}=10^2$.
The model with $\omega = 0.1$ shows the evolution qualitatively similar
to the model in \S \ref{subsec:rotation}, i.e., 
after the supercritical core formation, a bar mode 
grows in time without fragmentation.
On the other hand, when the rotation rate is doubled ($\omega=0.2$), 
the bar grows more rapidly even at the 
early time shown in panel (b), when the maximum
surface density reaches 10. By this time, the entire bar (the darkest
part in panel [b]) becomes magnetically supercritical.
By the time shown in panel (c) ($\Sigma _{\rm max}= 10^2$), 
the $m=2$ mode breaks up into two separate cores, each of 
which continues to collapse dynamically. (We checked that 
this model is stable to higher $m$ modes.) The subsequent 
evolution of each core is qualitatively similar to that of
the models shown in the previous section, i.e., the aspect ratio of
each core is more or less frozen at $\sim $ 2 during the early phase of
supercritical collapse, and increases rapidly at the very end of the
supercritical collapse.

For the separate core formation case, the magnetic field has a positive
effect on the fragmentation.  We calculated the models of a weaker 
and a stronger magnetic field for the same set of parameters $n=4$, 
$r_0=5\pi$, $\omega=0.2$, and $A=0.05$, 
and found that the cloud with a weaker magnetic field does not 
break up into two cores (it collapses into a single bar instead), 
whereas the cloud with a stronger magnetic field fragments into 
two cores at a lower density and with a wider separation. More 
centrally-condensed clouds with $n=2$ and $r_0 \lesssim 20\pi$ 
do not fragment into two cores even in the presence of a 
rapid rotation ($\omega \sim 0.3$).
These dependences on $\Gamma_0$ and $n$ are the same as those of
the multiple fragmentation case investigated by \citet{ZLi02}.
In this sense, the separate core case is similar to the multiple 
fragmentation case, although the bar fragmentation into two cores 
appears to depend more strongly on the rotation rate. 

At the time shown in panel (c), the separation of two cores is about 
a tenth of the initial cloud radius, which corresponds to $5\times 
10^4$ AU with the fiducial choice of $\Sigma _0 = 0.01$ g cm$^{-2}$ 
and $T_{\rm eff} = 30$ K. This separate core formation case may thus 
be responsible for the formation of wide binaries with separations 
of the order of $10^4$ AU, although substantial orbital evolution 
is likely after the initial fragmentation, since the cores are not 
kept apart by rotation.  

To summarize, when the initial rotation is relatively fast and the 
density distribution in the central region is relatively flat, the 
$m=2$ mode can break up into two separate cores in the isothermal 
collapse phase after the central region becomes supercritical (e.g., 
for $n=4$, $\omega \sim$ 0.2, $\Gamma_0 \sim 1.5$, and $r_0 
\gtrsim 5\pi$).
We note that this separate core formation differs from that found
previously for non-magnetic clouds, which requires a rapid rotation 
or highly unstable initial condition \citep[e.g.,][]{TTsuribe99}. 
In our case, the cloud is initially stable to gravitational collapse, 
and the fragmentation is induced by ambipolar diffusion. Indeed, 
we find that clouds with relatively weak magnetic fields do not 
fragment into separate cores during the isothermal collapse phase.

\subsection{Bar Fragmentation and Merging of Fragments}

According to linear perturbation theory, long filamentary clouds tend 
to break up gravitationally into fragments if their lengths exceed 
several times the diameter. This is indeed the case for the first
bars formed in relatively massive, centrally-condensed clouds
(with a relatively large $r_0$ and $n\lesssim 4$, 
e.g., for $n=2$, $\Gamma_0 \sim 1.5$, $\omega \sim 0.2$, 
and $r_0 \gtrsim 15\pi$), as illustrated in Fig. \ref{fig:7}, 
where the snapshots of a cloud with $n=2$, $\Gamma _0 = 1.5$, 
$r_0=15\pi$, and $\omega=0.2$ are shown at three different times.   
In this model, fragmentation starts when the bar becomes opaque 
to dust emission, which retards the radial contraction (at 
$\Sigma_c\gtrsim 10^5$).
At first, the bar produces two dense fragments near the center at 
$\Sigma \sim 10^5$, and then the fragmentation apparently 
propagates towards the ends of the bar.
This apparent propagation of
the fragmentation is due to the fact that the growth rate 
of the gravitational fragmentation is larger for the 
higher density region closer to the center.   
By the time shown in panel (a), there is a fragment formed at 
each of the two ends of the bar.  
This density peak formation at the end was first noticed 
by \citet{RLarson72} and subsequently investigated extensively 
by \citet{PBastien83} and \citet{IBonnell91}. It is caused by
a local minimum in the gravitational potential. 
The masses of the fragments formed by bar fragmentation are on the 
order of $10^{-2} M_\odot$. These masses agree with those 
expected from the linear theory of fragmentation of cylindrical
clouds \citep[e.g.,][]{RLarson85, FNakamura93}. 
By the time shown 
in panel (b), the two pairs of fragments near the center have 
merged to produce one, more massive, pair. This process of 
successive bar fragmentation and subsequent merging is repeated, 
as shown in panel (c), where the two most massive fragments 
are about to merge at the center. We will discuss further the 
evolution of this multiple fragment system in \S \ref{sec:discussion}.

We note that bar fragmentation was also observed in the numerical 
simulations of non-magnetic cloud collapse \citep[e.g.,][]{ABoss93,
MBate97}.   
The non-magnetic bars are typically much shorter than the one presented 
in this subsection. To 
form such a long bar in the non-magnetic case, the initial cloud must 
contain many thermal Jeans masses and thus be highly unstable. In our 
case, the initial cloud is stablized by a strong magnetic field, 
whose support weakens with time as a result of ambipolar diffusion. 
\citet{FNakamura97} studied bar formation in clouds with frozen-in 
magnetic fields. They found bars that are prone to gravitational 
breakup according to the linear theory of fragmentation. These bars 
are however shorter than the one discussed here. They are expected 
to produce a smaller number of fragments. 

\subsection{Disk Formation and Fragmentation}

When the initial clouds are centrally condensed and/or less massive
(e.g., for $n=2$, $\omega \sim0.1$, 
$\Gamma_0 \sim 1.5$, and $r_0 \lesssim 15\pi$), 
the length of the bar does not exceed the critical wavelength
for gravitational fragmentation by the time of first bar formation.
It would be difficult for such a short bar to break up into blobs 
gravitationally. In this case, a rapidly-rotating disk is formed 
in the presence of slow rotation.
Here, we show an example of the rotating disk formation.
Figure \ref{fig:8} shows the snapshots at six different times 
of a model with $n=2$, $\Gamma _0 = 1.5$, $r_0=5\pi$, and 
$\omega=0.1$, which is identical to the slowly rotating model
discussed in \S \ref{subsec:rotation}, except for a smaller 
characteristic radius $r_0$ ($5\pi$ instead of $10\pi$). 
For comparison, the linear density contours of the central disk is
depicted in Fig. \ref{fig:9} at three selected times. In this 
model, the bar mode does not grow much by the end of isothermal 
collapse because of the relatively small cloud radius $r_0$\footnote{ 
We estimate that the central plateau region of this cloud has a 
size roughly $2-3$ times the effective Jeans length, judging from 
the fact that, for $n=2$, reference clouds with $r_0 \lesssim 
2\pi$ expand rather than contract.}. 
Because the length of the bar is smaller than the critical wavelength 
for gravitational fragmentation, it does not fragment. 
Instead, a round dense region, which we called the ``truly first 
core'', is formed at the center due to a supersonic infall motion 
along the bar.  

Initially, the truly first core is supported primarily by the thermal 
pressure.  As ambient gas with a larger angular
momentum accretes onto it, the rotational support becomes dominant.
As a result, the central core evolves into a rapidly-rotating,
opaque disk, which we will sometimes refer to as the ``first disk''. 
The interaction between the disk and the large-scale bar mode 
outside the disk excites another bar-mode growing inside the disk.  
We believe that the secondary bar growth is due to rotational instability 
because the ratio of the rotational energy and the gravitational energy 
of the first disk reaches the critical value of $\beta_{\rm cr} \sim 0.3$
\citep[e.g.,][]{RDurisen86,MBate98,JTohline02}.
By the time shown in panel (c) of Fig. \ref{fig:9}, the bar breaks up
into two blobs, which appear to be self-gravitating.   We stopped our 
calculation at this epoch of fragmentation because the volume densities 
of the blobs reach the critical density beyond which H$_2$ dissociation 
starts and the equation of state of the gas is expected to change again.  

The fate of the two blobs formed in the disk is uncertain. They may merge 
together during the subsequent evolution.  Since the rotation period of
the disk is faster than that of the large-scale bar outside the disk, the 
interaction between the two bar modes generates several blobs in the 
disk, some of which appear to be self-gravitating.  These blobs 
are expected to interact gravitationally and whether any of them can
survive the interaction remains to be determined. This type of 
fragmentation and merging in a rotating disk was also observed in the 
numerical simulations of non-magnetic cloud collapse \citep[e.g.,][]
{IBonnell94, AWhitworth95,ABurkert97,TMatsumoto02}.

\section{Discussion}
\label{sec:discussion}

\subsection{Fragmentation of Magnetic Clouds}

In the previous sections, we have shown that core-forming clouds 
tend to become unstable to 
the bar mode after the supercritical core formation. The numerical 
results presented in this paper and in Li \& Nakamura (2002) 
indicate that fragmentation of magnetic clouds happens 
readily over a wide range of initial cloud parameters. On the
surface, this result appears to contradict earlier numerical 
studies of magnetic cloud fragmentation, which suggest that 
strong magnetic fields tend to stifle fragmentation of dynamically 
collapsing clouds \citep{EDorfi82,WBenz84,GPhillips86a,GPhillips86b}.
For example, \citet{GPhillips86a} performed three-dimensional 
MHD simulations on initially magnetically supercritical clouds and 
found that none of his clouds show any evidence for fragmentation
even in the presence of a large density perturbation with a fractional
amplitude of 50\%. We emphasize that there are two important 
differences between our model and the previous ones. First, previous 
studies considered a frozen-in field, whereas in our study ambipolar 
diffusion is a crucial ingredient.
Our initial cloud, which is supported by a strong magnetic field, 
is stable to gravitational fragmentation as well as dynamic collapse.
The weakening of the magnetic support driven by ambipolar diffusion 
is what initiates the dynamic collapse and fragmentation. 
Second, the contribution of the magnetic tension to the cloud support 
is much weaker in the earlier studies; indeed, most of them took a uniform 
magnetic field as the initial state (which of course has no magnetic 
tension). The beneficial effect of magnetic tension on cloud 
fragmentation was discussed by \citet{ABoss00, ABoss02}, who showed 
that a strong magnetic tension can slow down cloud contraction 
appreciably, which tends to enhance cloud fragmentation.

Recently, \citet{JTohline02} reviewed the theoretical studies on binary
star formation, concentrating on the non-magnetic case.
He found broad agreement among numerical simulations by various groups
that non-magnetic clouds are susceptible to prompt fragmentation if they
collapse from an initial configuration that is relatively uniform in
density and contains more than a few Jeans masses of material. If
supported only by thermal pressure, such a configuration is highly 
unstable. The question is then \citep{JTohline02}: how is nature 
able to produce such an unstable configuration in the first place? 
We believe that the key to resolve this 
apparent difficulty is the presence of a strong magnetic field, which 
stabilizes the entire cloud even when the cloud mass is well above one 
thermal Jeans mass \footnote{The cloud can also be supported by
turbulence, although the details of cloud evolution driven by
turbulence decay remain to be worked out \citep{PMyers99}.} 
\citep[e.g.,][]{SLizano89}. Specific models of core formation
in  magnetically subcritical clouds appear able to explain the mass
distribution and extended subsonic infall motions inferred in several
starless cores \citep[e.g.,][]{CLee01,GCiolek00,RInde00} and are 
consistent with pattern of chemical differentiation observed in 
L1544 \citep{ZLi02b}, provided that the initial flux-to-mass ratio lies 
within a factor of two or so of the critical value, as adopted in 
this paper. 

\subsection{Elongation of Dense Cloud Cores}

Dense cores of molecular clouds are observed to have significant 
elongation, with a typical aspect ratio of 2
in the plane of sky \citep{PMyers91}. Their true three-dimensional shapes
are probably triaxial, as inferred statistically by \citet{CJones01} and 
\citet{SGoodwin02}. We have previously noted that the triaxial core shape 
can naturally result from nonlinear growth of the bar mode in the direction 
perpendicular to the field lines, coupled with the flattening of the
mass distribution along the field lines \citep{FNakamura02}. 
The models presented in 
\S \ref{sec:results} strengthen this conclusion. We find that the 
aspect ratio of the bar 
is more or less independent of the initial model parameters. 
It stays at about 1.5 $-$ 2 during the period when $\Gamma$ 
decreases from $\sim 1$ to $\sim 0.5$, after the core has become 
supercritical (which makes bar growth possible) but before the 
ratio of effective gravity to effective pressure force exceeds 
$\sim 2$ (when the bar elongation is greatly amplified through 
the Lin-Mestel-Shu instability). This initial period of 
supercritical phase is also the phase most likely observed as a 
starless core, when the core is dense enough to be identified 
as a core using high density tracers (such as NH$_3$ or CS), but 
before rapid collapse sets in which quickly produces a stellar 
object at the center, ending the starless phase.

In this supercritical collapse phase, the decrease of the central 
flux-to-mass ratio saturates at about half the critical value, 
almost independent of the initial parameters (see panel [b] of 
Fig. \ref{fig:2}). This saturation of flux-to-mass ratio is also 
observed in axisymmetric models \citep{SBasu95}. It means that 
the supercritical cores remain significantly magnetized, and 
magnetic fields may not be completely ignored even during the
dynamic collapse phase of star formation. 

\subsection{Three Types of Binary and Multiple Star Formation}
\label{types}

Observations of nearby star forming regions have established that 
binary fraction among pre-main sequence stars is at least as 
high as among the main-sequence stars. In some nearby low-mass star 
forming regions like Taurus-Auriga, the binary frequency is even 
higher than that of main-sequence stars. 
In such low-mass star forming regions, stellar densities 
($n_* \sim 10$ pc$^{-3}$) are much too low to allow the formation 
of binaries by capture or their destruction 
by gravitational encounters within their ages of $\sim$ 1 Myr.
This indicates that the observed high binary frequencies more or less 
reflect the pristine population of proto-binaries 
at the epoch of formation, suggesting that binary and small multiple 
stars are formed prior to pre-main sequence phase by gravitational
fragmentation of parent dense molecular cloud cores.
In fact, recent high resolution millimeter observations of dust continuum
have revealed evidence of gravitational fragmentation of protostellar
cores in the earliest, Class 0 and I phases. Such observations were
summarized in \citet{LLooney00}, who classified young multiple star 
systems into three types: 
(1) separate envelope (separation $\ge 6000$ AU), (2) common
envelope (separation 150 $-$ 3000 AU), and (3) common disk systems
(separation $\le 100$ AU).
Separate envelope systems exhibit clearly distinct centers of
gravitationally bound cores with separations greater than 6000 AU and the
components are surrounded by low density material
(e.g., NGC1333 IRS2, SVS13).
Common envelope systems have one primary core of gravitational
concentration which apparently breaks up into multiple fragments with
separations of 100 $-$ 3000 AU (e.g., IRAS 16293-2422).  
Common disk systems have circumbinary disk-like envelopes 
surrounding young stellar objects with separations smaller than 100 AU 
(e.g., L1551 IRS5).

We have also uncovered three distinct types of fragmentation of magnetic 
clouds in our numerical experiments. They correspond roughly with the 
above empirical classification of young binary and multiple star systems.
Based upon the results presented in \S \ref{sec:fragmentation}, we 
classify theoretically the process of fragmentation into three types, 
depending on the initial model parameters:
(1) {\it separate core formation}, in which the cloud breaks up 
into two separate cores during the isothermal collapse phase, 
(2) {\it bar fragmentation}, in which 
the aspect ratio of the bar becomes large enough to break up 
into multiple fragments along the major axis and the fragmentation 
takes place after the bar has become opaque to dust emission, 
and (3) {\it disk fragmentation}, in which the 
central region, after becoming opaque to dust emission, 
evolves into a rapidly rotating disk, which fragments into 
multiple blobs. These cases are discussed in turn. 

When the initial cloud has a relatively flat central part and is
rotating relatively rapidly (e.g., $\omega \sim 0.2$ for $n\sim 4$), 
it fragments into two separate cores 
during the isothermal collapse phase in the presence of an $m=2$ 
perturbation after the central region becomes magnetically 
supercritical ({\it separate core formation}). Morphologically, 
this case resembles the separate envelope systems discussed in 
\citet{LLooney00}. 
The early phase of fragmentation also looks like the double-core 
in the extinction map of Fest 1-457 \citep{JAlves02} (see 
Laundardt [2001] for another example of
double-core, in Bok globule CB 230). When first produced, the 
dense cores have insufficient orbital rotation to prevent them 
from spiraling towards each other. We expect their separation 
to shrink appreciably with time, producing perhaps a highly 
eccentric system. Whether they will eventually merge together 
is unclear at the present. We suspect that the cores 
will survive to become individual star/stellar system, since 
the density is much higher inside the cores than outside by
the end of our calculations. This means that it takes the cores 
much less time to collapse onto themselves 
to form stars than to merge together. Moreover, there
exists a strong magnetic field outside the cores which can prevent 
direct core collision. In our separate core formation case, 
the density of the central inter-core region decreases with 
time as matter falls towards the cores, whereas the field strength
increases due to magnetic flux diffusion out of the cores into 
the inter-core region. Indeed, the central inter-core region may 
become magnetically subcritical again at the later stages of 
evolution. This region serves as a ``magnetic cushion'' to direct core 
collision. We propose that the separate core formation is a mechanism 
for producing wide binaries with separations of $\sim 10^4$ AU.
Each core will have a second chance at fragmentation at higher 
densities, when a first bar or rotating disk is formed. The secondary 
fragmentation within each core could lead to the formation of a small 
(hierarchical) multiple star system. As mentioned earlier, separate 
core formation in the presence of an $m=2$ perturbation has much in 
common with the multiple fragmentation induced by higher order 
perturbations that we studied previously in \citet{ZLi02}. Multiple 
core formation is an related channel for wide binary and multiple star 
formation. 

When the initial cloud contains a mass well above the thermal Jeans mass
and has a moderately centrally condensed density distribution
(e.g, $r_0 \gtrsim 15\pi$ for $n \sim 2$ and $\omega \sim 0.1-0.2$), a long 
bar is formed, which can break up into several (pairs of) fragments 
after the bar becomes opaque to the dust emission ({\it bar 
fragmentation}).  The multiple fragments grow in mass due to both merging 
and accretion. This bar fragmentation case would correspond to the common 
envelope systems of \citet{LLooney00}, since the whole bar is embedded 
within a common core out of which the bar is formed. The bar fragmentation 
occurs at a density of $\sim 10^{12}$ cm$^{-3}$.  According to linear 
perturbation theory, a highly elongated bar breaks up gravitationally 
into fragments, with separations comparable to the wavelength of the most 
unstable linear perturbation, which corresponds to $\sim 10^2$ AU at
$\sim 10^{12}$ cm$^{-3}$. The typical masses of the fragments are 
in the substellar regime. 
Some of the fragments could be ejected 
out of the system in the course of dynamical interaction
\citep[see also][]{MBate02}. Those ejected 
may become isolated brown dwarfs \citep{BReipurth01} and
perhaps even free-floating massive planets. Interestingly, the cloud 
parameters derived by \citet{GCiolek00} for the well-studied starless 
core L1544 put the cloud in the bar fragmentation category. Therefore, 
L1544 appears to be a progenitor of a binary or small multiple star 
system of the common envelope type. 

When the initial cloud mass in the central plateau region is comparable
to the thermal Jeans mass (e.g., $r_0\lesssim 15\pi$ for $n\sim 2$
and $\omega \sim 0.1$), 
the bar mode does not grow much during the
isothermal collapse, and it is difficult for the (relatively short) bar 
to break up into fragments ({\it disk fragmentation}).
In this case, the central region evolves into a rapidly rotating disk
whose size depends on the rate of initial rotation.
The interaction of the central disk with the bar generates complex
structures in the disk where a secondary bar mode is excited by rotational
instability if $\beta \gtrsim 0.3$. Typically, the size of centrifugal 
radius is of the order of 10 AU. When the disk fragments into blobs, the 
separations of the blobs should be of the same order of the disk radius.
This case may be responsible for the formation of binary stars with
a separation smaller than $10^2$ AU. 
This case would be related to the 
common disk system of \citet{LLooney00}.

As mentioned in \S~\ref{sec:method}, we did not consider angular momentum 
transfer in the vertical direction, which can brake the cloud rotation 
efficiently during the relatively long magnetically subcritical phase 
of evolution \citep{TNakano72,SBasu94}. Magnetic braking is thought to 
be responsible for the low rates of rotation observed in dense cores 
\citep{AGoodman93} and adopted in 
this paper. Once the central region of the cloud becomes supercritical, 
the peak density increases in a runaway fashion, leaving little time 
for the magnetic braking to operate efficiently \citep{SBasu94}. We 
therefore expect the formation of separate cores, which occurs in the 
runaway collapse phase, to be little affected by magnetic braking. After 
a bar or a disk becomes dense enough to be opaque to dust emission, it 
can be supported by thermal pressure for a time long compared to the 
local dynamic time, and magnetic braking can in principle become 
efficient again \citep{RKrasnopolsky02}. However, at such high densities, 
magnetic fields start to decouple from matter \citep{RNishi91}, which 
could render the braking inefficient. We will postpone a detailed 
treatment of the magnetic braking including the effects of decoupling 
to a future work. 

\section{Conclusion} 
\label{sec:conclusion}

In this paper, we have followed the early phases of cloud fragmentation 
in magnetic clouds, using a two-dimensional MHD code based on the 
thin-disk approximation.
At the end of our calculations, we obtained dense,
discrete fragments out of initially axisymmetric clouds perturbed 
by an $m=2$ mode of modest fractional amplitude $A=0.05$. The masses 
of the fragments are still small, on the order of $10^{-2}-10^{-1}$ 
M$_\odot$. We expect the fragments to grow to stellar masses, by 
merging and further accretion, except perhaps those ejected dynamically 
out of the system soon after the fragmentation. The masses of those
ejected may remain frozen in the substellar regime. 

\begin{figure}
%\epsscale{0.6}
\plotone{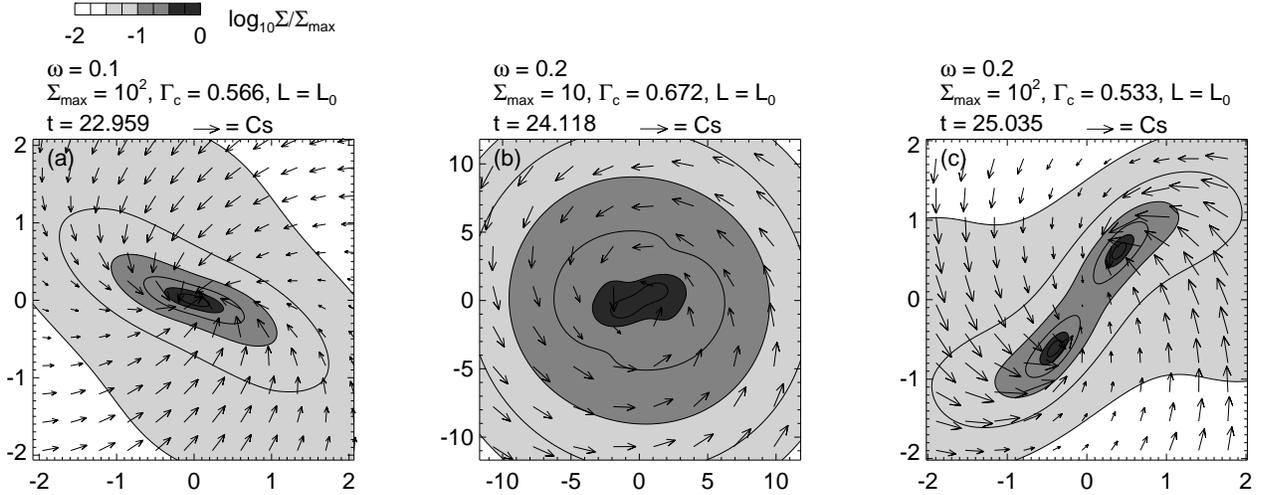} 
\caption{Snapshots of the surface density distribution and velocity
field of two models (a) with $(r_0, n, \omega, \Gamma _0, A)=
(\underline{5\pi}, \underline{4}, \underline{0.1}, 1.5, 0.05)$,
and (b) and (c) with $(r_0, n, \omega, \Gamma _0, A)=
(\underline{5\pi}, \underline{4}, \underline{0.2}, 1.5, 0.05)$.
The contours, arrows and notations have the same meaning as in 
Fig.~\ref{fig:1}.
The parameters having different values from those of Fig. \ref{fig:1}
are underlined. In the more rapidly rotating model, 
the initially-imposed $m=2$ mode evolves into a long bar, which
breaks up into two distinct cores (panel c).
\label{fig:6}}
\end{figure}

\begin{figure}
%\epsscale{0.6}
\epsscale{1.0}
\plotone{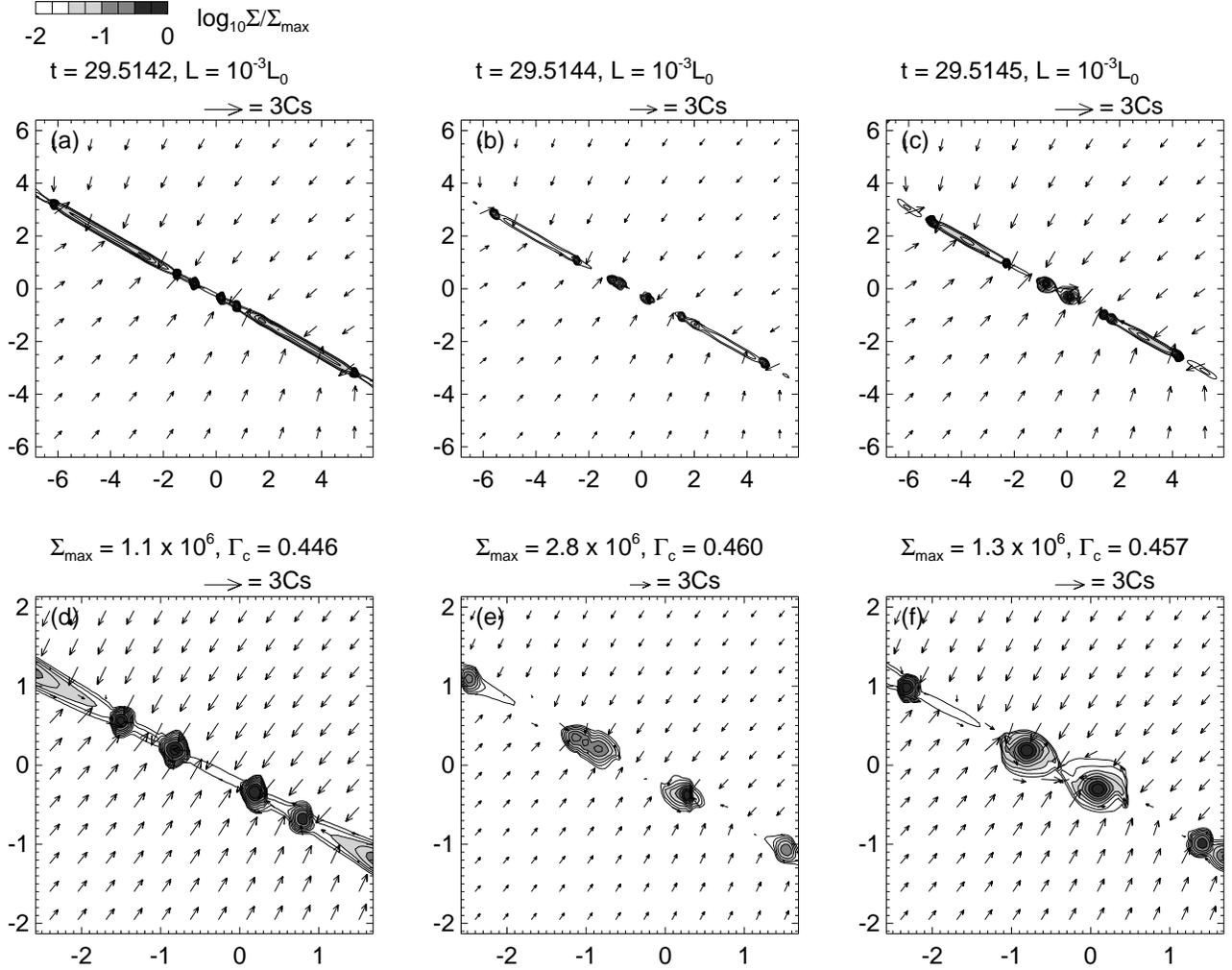}
\caption{Snapshots of the surface density distribution and velocity
field of the model with $(r_0, n, \omega, \Gamma _0, A)=
(\underline{15\pi}, 2, \underline{0.2}, 1.5, 0.05)$
at three different times: (a) $t = 29.5142$, 
(b) $t = 29.5144$, and (c) $t = 29.5145$.
Panels (d) though (f) show the enlargement of the central region of
 panels (a) through (c), respectively.
The contours, arrows and notations have the same meaning as in 
Fig.~\ref{fig:1}.
The parameters having different values from those of Fig. \ref{fig:1}
are underlined.
In this model, the aspect ratio of the first bar becomes very large
and the fragmentation starts near the center and at the ends (panel a).
Subsequent fragmentation occurs successively in the high
 density central region, and seemingly propagates towards the
ends (panel b). 
At the same time, the central fragments merge together to form 
more massive fragments (panel b).  
This process is repeated at the time shown in panel (c).
\label{fig:7}}
\end{figure}

\begin{figure}
%\epsscale{0.6}
\epsscale{1.0}
\plotone{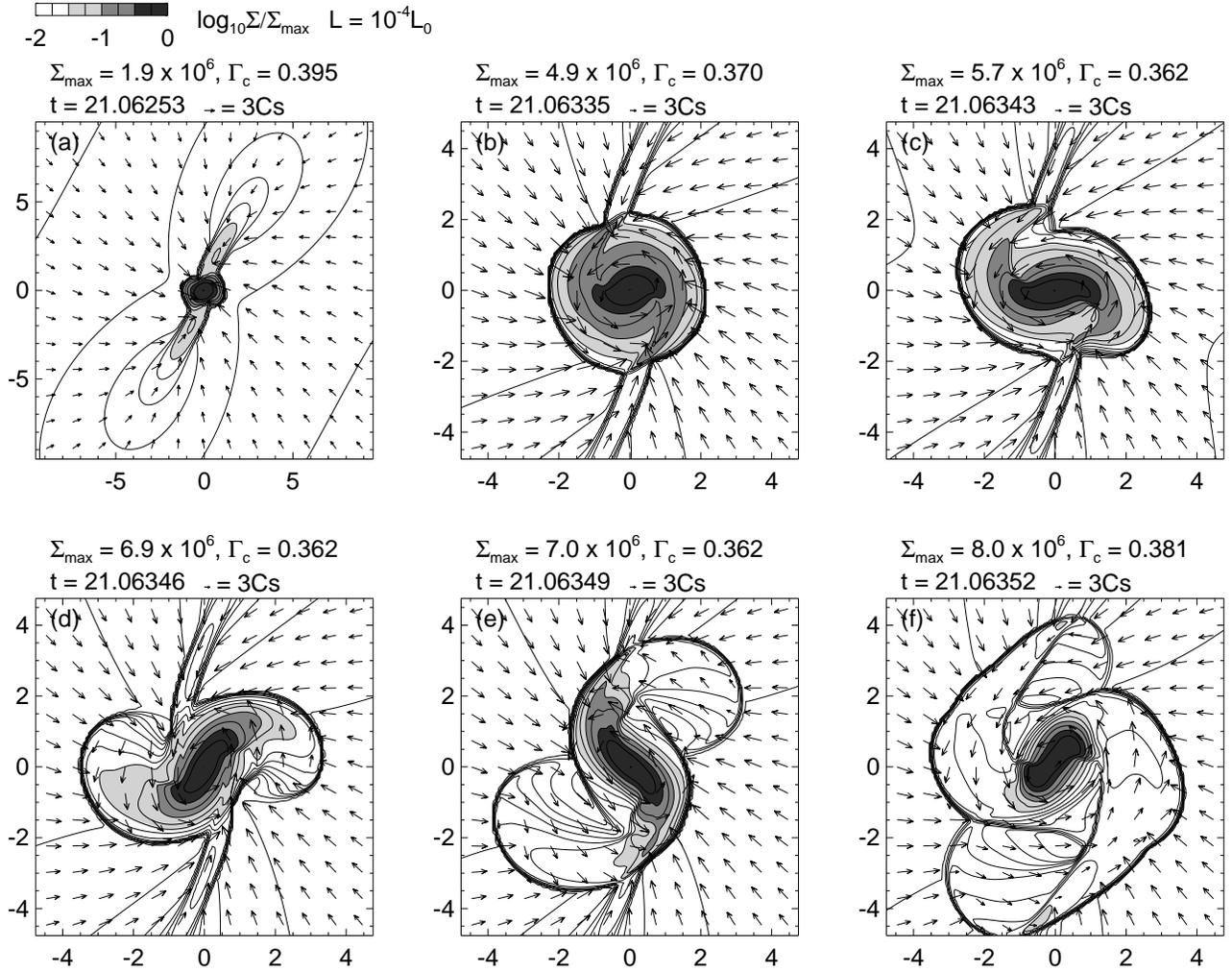}
\caption{Snapshots of the surface density distribution and velocity
field of the model with $(r_0, n, \omega, \Gamma _0, A)=
(\underline{5\pi}, 2, \underline{0.1}, 1.5, 0.05)$
at six different times.  
The contours, arrows and notations have the same meaning as in 
Fig.~\ref{fig:1}.
The parameters having different values from those of Fig. \ref{fig:1}
are underlined.
As the infall along the major axis becomes significant, 
the truly first core is formed at the center, which evolves into
a rapidly rotating disk (panel a).  
In the rotating disk, an $m=2$ mode develops, which  
interacts with the inflows along the first bar (panels b and c),
generating shock waves in several places (panels d and e).  
The disk fragments into several blobs due to complex shock interaction 
(panels e and f).
\label{fig:8}}
\end{figure}

\begin{figure}
%\epsscale{0.6}
\epsscale{1.0}
\plotone{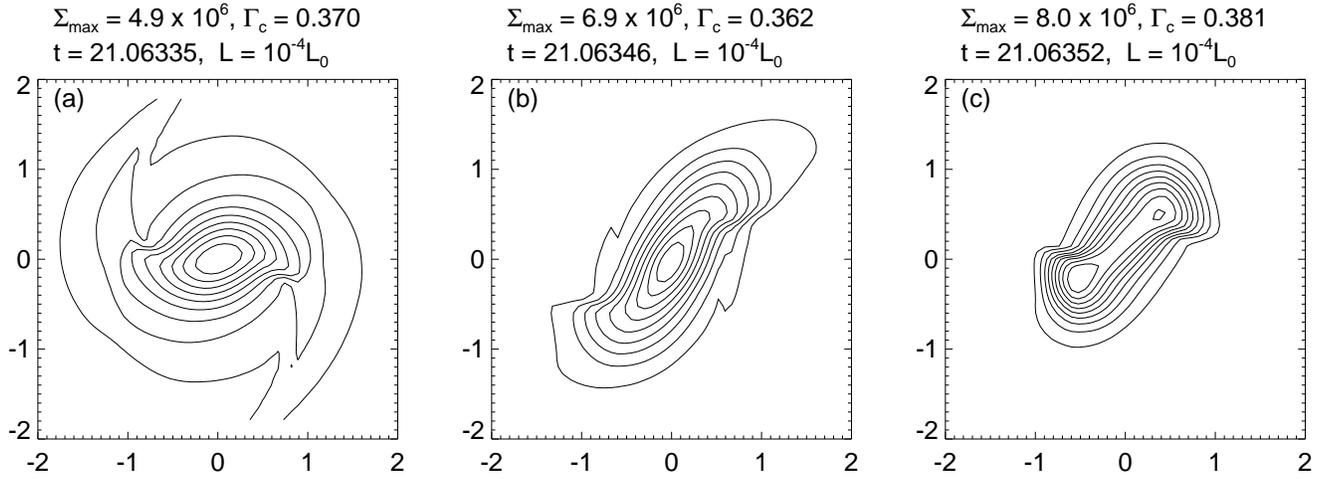}
\caption{Surface density distributions of the central rotating disk for 
the model with $(r_0, n, \omega, \Gamma _0, A)=
(\underline{5\pi}, 2, \underline{0.1}, 1.5, 0.05)$
at three different times. 
Panels (a), (b) and (c) are at the same times as panels (b), (d), and (f)
 of Fig. \ref{fig:8}. The surface density contours are linear in scale,
plotted at levels of 0.1, 0.2, $\dots$, $0.9$ of the maximum, 
$\Sigma_{\rm max}$. At the time shown in panel (c), the bar breaks 
up into two blobs, which appear to be self-gravitating.
\label{fig:9}}
\end{figure}

We found three types of fragmentation in magnetic clouds. They are (1)
a cloud breaking up during the isothermal phase of cloud evolution 
into separate cores, each of which collapses more or less independently 
into a star/stellar system (panels [b] and [c] of Fig. \ref{fig:6}), 
(2) a core collapsing into a needle-like, 
opaque ``first bar'', which breaks up into several 
fragments (Fig. \ref{fig:7}), and (3) 
a rapidly rotating, opaque ``first disk'' producing self-gravitating 
blobs, which may survive to become seeds of relatively close binary 
and multiple systems (Figs. \ref{fig:8} and \ref{fig:9}).
The fragments in the above cases share, 
respectively, a common cloud, common core, and common disk. They 
correspond loosely to the empirical classification of 
independent envelope, common envelope and common disk systems for the
youngest, embedded binary and multiple systems. Much work is needed to 
firm up this connection.

In the near future, we plan to extend our numerical simulations beyond
the initial fragmentation to the protostellar accretion phase, to 
determine the physical properties of the binary and multiple systems 
formed in magnetic clouds, such as their eccentricity, mass ratio, and 
orbital period. In a longer run, we hope to generalize the calculations
to 3D, which will allow for a detailed study of magnetic braking, which 
may have a profound impact on the final orbits of the formed stellar 
system. 

\acknowledgments

Numerical computations in this work were carried out  at the Yukawa 
Institute Computer Facilities, Kyoto University. F.N. gratefully 
acknowledges the support of the JSPS Postdoctoral Fellowships for 
Research Abroad, and Z.Y.L. is supported in part by NASA.

\begin{deluxetable}{lllllll}
%\scriptsize
%\rotate
\tablecolumns{8}
\tablecaption{Parameters of Models \label{table1}}
\tablewidth{5.5in}
\tablehead{ 
  \colhead{$r_0$}     & \colhead{$n$}          & \colhead{$\omega$}
 &\colhead{$\Gamma_0$}  & \colhead{perturbation} & \colhead{$A$}
 &\colhead{display}
}
\startdata
$10\pi$  & 2 & 0.0 & 1.5 & $m=2$  & 0.05 & Fig. \ref{fig:1} (standard model)\\
$10\pi$  & 2 & 0.0 & 1.25 & $m=2$  & 0.05 & Fig. \ref{fig:4}a\\
$10\pi$  & 2 & 0.1 & 1.5 & $m=2$  & 0.05 & Fig. \ref{fig:5}\\
$15\pi$  & 2 & 0.0 & 1.5 & $m=2$ & 0.05 & Fig. \ref{fig:4}b\\
$15\pi$  & 2 & 0.2 & 1.5 & $m=2$ & 0.05 & Fig. \ref{fig:7} (bar fragmentation)\\
$5\pi$  & 2 & 0.1 & 1.5 & $m=2$ & 0.05 & Figs. \ref{fig:8}, \ref{fig:9}
 (disk fragmentation)\\
$ 5\pi$  & 4 & 0.0 & 1.5 & $m=2$ & 0.05 & Fig. \ref{fig:4}c\\
$ 5\pi$  & 4 & 0.1 & 1.5 & $m=2$ & 0.05 & Fig. \ref{fig:6}a\\
$ 5\pi$  & 4 & 0.2 & 1.5 & $m=2$ & 0.05 & Figs. \ref{fig:6}b,c 
(separate core formation)\\
\enddata
\tablecomments{All models are computed on an initial grid of $512\times 
512$ in the disk ($x$-$y$) plane. For the standard model shown in 
Fig. \ref{fig:1}, the grid number is increased to $1024\times 1024$ 
for each level of refinement.
}
\label{tab:1}
\end{deluxetable}

%\clearpage

\end{document}